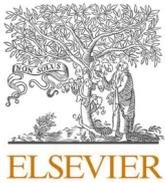
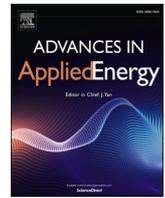

# Physics-informed machine learning for building performance simulation-A review of a nascent field

Zixin Jiang [a], Xuezheng Wang [a], Han Li [b], Tianzhen Hong [b], Fengqi You [c], Ján Drgoňa [d], Draguna Vrabie [e], Bing Dong [a,*]

[a] *Department of Mechanical & Aerospace Engineering, Syracuse University, 263 Link Hall, Syracuse, NY 13244, United States*
[b] *Building Technology & Urban Systems Division, Lawrence Berkeley National Laboratory, Berkeley, 94720, CA, USA*
[c] *Cornell University, Robert Frederick Smith School of Chemical and Biomolecular Engineering, 120 Olin Hall, Ithaca, NY 14853, USA*
[d] *Department of Civil and Systems Engineering and the Ralph S. O'Connor Sustainable Energy Institute (ROSEI) at Johns Hopkins University (JHU), MD 21218, USA*
[e] *Pacific Northwest National Laboratory, Richland, WA 99354, USA*

ABSTRACT

Building performance simulation (BPS) is critical for understanding building dynamics and behavior, analyzing the performance of the built environment, optimizing energy efficiency, improving demand flexibility, and enhancing building resilience. However, conducting BPS is not trivial. Traditional BPS relies on accurate building energy models, which are primarily physics-based and heavily dependent on detailed building information, expert knowledge, and case-by-case model calibrations, significantly limiting their scalability. With the development of sensing technology and the increased availability of data, there is growing attention and interest in data-driven BPS. However, purely data-driven models often suffer from limited generalization ability and a lack of physical consistency, resulting in poor performance in real-world applications. To address these limitations, recent studies have begun integrating physics priors into data-driven models, a methodology known as physics-informed machine learning (PIML). PIML is an emerging field where its definitions, methodologies, evaluation criteria, application scenarios, and future directions remain open. To bridge those gaps, this study systematically reviews the state-of-the-art PIML for BPS, offering a comprehensive definition of PIML and comparing it to traditional BPS approaches regarding data requirements, modeling effort, performance, and computational cost. We also summarize the commonly used methodologies, validation approaches, application domains, available data sources, open-source packages, and testbeds. In addition, this study provides a general guideline for selecting appropriate PIML models based on BPS applications. Finally, this study identifies key challenges and outlines future research directions, providing a solid foundation and valuable insights to advance R&D of PIML in BPS.

## Nomenclature

**Table 1**
Nomenclature table of terms and acronyms used in the paper.

| Notation | Meaning | Notation | Meaning |
|---|---|---|---|
| BPS | Building Performance Simulation | PIML | Physics-Informed Machine Learning |
| R&D | Research & Development | ML | Machine learning |
| HVAC | Heating, Ventilation, and Air Conditioning | CFD | Computational Fluid Dynamics |
| IEQ | Indoor Environment Quality | BTM | Behind-The-Meter |
| DERs | Distributed Energy Resources | FDD | Fault Detection and Diagnostics |
| IAQ | Indoor Air Quality | TRV | Temperature Response Violation |
| ODE | Ordinary Differential Equation | RC model | Reduced Resistors and Capacitors Model |

*(continued on next page)*

Table 1 *(continued)*

**Table 1** (*continued*)

| Notation | Meaning | Notation | Meaning |
|---|---|---|---|
| NN | Neural Network | CNN | Convolutional Neural Network |
| ANN | Artificial Neural Network | LSTM | Long Short-Term Memory |
| RNN | Recurrent Neural Network | GNN | Graph Neural Network |
| GP | Gaussian Process | DBN | Diagnostic Bayesian Network |
| ARMA | Auto Regressive Moving Average | $\mathcal{F}$ | Latent Representation |
| $x$ | State Input | $\tilde{x}$ | State Estimation |
| MSE | Mean Squared Error | PINN | Physics-Informed Neural Network |
| MAE | Mean Absolute Error | MAPE | Mean Absolute Percentage Error |
| RMSE | Root Mean Squared Error | MMD | Maximum Mean Discrepancy |
| AD | Automatic Differentiation | RTE | Radiative Transfer Equation |
| PSO | Particle Swarm Optimization | MPC | Model Predictive Control |
| DPC | Differentiable Predictive Control | RL | Reinforcement Learning |
| PDEs | Partial Differential Equations | RANS | Reynolds-averaged Navier–Stokes |

## 1. Introduction

Buildings significantly contribute to global energy consumption and carbon emissions, accounting for approximately 34 % of global energy use and 37 % of $CO_2$ emissions [1]. Optimizing building energy efficiency is therefore critical for mitigating global warming, addressing the energy crisis, and moving toward carbon neutrality [2]. Building performance simulation, also called building simulation [3] or building energy modeling, plays a pivotal role in this endeavor.

BPS is a computational technique that uses mathematical models to simulate and predict building energy flows, airflows, lighting, thermal comfort, and other indoor environmental quality (IEQ) metrics based on building characteristics, systems, and operations such as architectural design, floor plans, envelopes, lighting systems, Heating, Ventilation, and Air Conditioning (HVAC) systems, plug load equipment, occupancy schedules, and local climate [4,5]. BPS has been increasingly utilized in building energy system simulation [6,7], building design [8], building environment assessment [9], building control operations [10,11], building-to-grid integration [12,13], building energy retrofitting [14], and urban energy planning [15]. Moreover, it can assist building owners, architects, engineers, and policymakers in understanding how buildings can be optimally designed, constructed, retrofitted, maintained, and operated [16].

In general, commonly used BPS methods can be classified into two categories: 1) physics-based (white-box) models and 2) data-driven (black-box) models [17]. Examples of physics-based models include EnergyPlus [18] and TRNSYS [19] for building energy simulation, Fluent [20] and CONTAM [21] for air-fluid simulation, Radiance [22] and Ecotect [23] for lighting simulation, and Odeon [24] for acoustic simulation. These models rely on detailed building metadata and solve governing equations—such as those for mass, momentum, and energy balance—based on fundamental physical principles. They are known for their high accuracy and ability to simulate system behavior under previously unobserved conditions [17]. However, three major barriers limit the large-scale application of physics-based models [16]: 1) **Extensive data requirements**—physics-based models typically require detailed information about building physical properties, such as geometry, envelope materials, equipment performance, occupancy schedules, and boundary conditions, which are often difficult to obtain in real-world applications; 2) **Significant modeling effort**—for building energy models, numerous parameters must be carefully calibrated on a case-by-case basis [17], while for finite element-based models, developing accurate geometry and generating meshes remain complex tasks [25]; and 3) **Substantial computational burden**—computation time increases exponentially from hours to days or weeks [26] with model complexity in terms of urban-scale energy modeling [27] and the fine-grained computational fluid dynamics (CFD) simulations involve solving high-dimensional sparse matrices, which remain prohibitively expensive, especially for practical turbulent systems involving multi-physical processes [28].

In contrast, data-driven models are developed purely based on mathematical mappings between system inputs and outputs, without requiring prior knowledge of the building's physical characteristics [29]. These models are primarily applicable after the building is constructed and less relevant at the design stage, when such models need to be trained with data from physics-based simulations. The typical pipeline for developing a data-driven model involves data collection from sensors or historical records, followed by data preprocessing, cleaning, and feature engineering. A suitable model is then selected for training by solving an optimization problem to minimize the error between the predicted outputs and the labeled data. After training and validation, the finalized model can be applied to various tasks. With the rapid development of sensing technology and advances in machine learning algorithms, these models have demonstrated higher accuracy, lower engineering costs, and reduced domain knowledge requirements compared to traditional physics-based models [29]. However, data-driven models also have four major limitations: 1) **Sensitivity to data quality and quantity** [30], as missing, incorrect, or biased data often lead to poor model performance; 2) **Lack of interoperability**, making it challenging to exchange, analyze, and interpret these models [31]; 3) **Limited generalization ability**, which makes it difficult to infer outputs for unseen conditions [32]; 4) **Absence of a guarantee for physical consistency**. An accurate predictive model does not necessarily ensure a well-behaved response model, and this mismatch can lead to significant failures in real-world applications [16,32].

To overcome the above-mentioned limitations of both physics-based and data-driven models in addressing complex engineering problems in large-scale real-world applications, recent research has explored their integration, commonly referred to as PIML [25]. The origins of this approach can be traced back to the early 1990s, when Dissanayake and Phan-Thien [33] implemented neural network-based functions to solve partial differential equations. With the increase in computational power and the availability of open-source packages such as PyTorch, more recent contributions by Kondor and Trivedi [34], and Mallat [35] further extended the concepts developed by Raissi et al. [36]., and introduced novel approaches such as discrete time-stepping schemes that allow PIML frameworks to be seamlessly integrated with any differential problem. These innovations significantly enhanced data efficiency, reduced computational costs, and attracted researchers seeking neural network-based solutions to complex problems. Since then, the scope of PIML has rapidly expanded, and it has been successfully applied in various fields, including but not limited to fluid mechanics [28], weather and climate modelling [37], structural analysis [38], anomaly detection and monitoring [39], etc.

In the context of BPS, PIML serves as a transformative approach. By combining the strengths of physics-based and data-driven models, PIML offers a robust alternative, enabling scalable, accurate, and physically consistent modeling for various building applications, including energy prediction and analysis, control and system optimization, IEQ assessment (thermal comfort, indoor air quality (IAQ), daylight, acoustics), airflow simulation, and behind-the-meter (BTM) distributed energy resources (DERs) integration. Furthermore, PIML has demonstrated higher accuracy, improved generalization ability, greater scalability, physical consistency guarantees, enhanced explainability, and reduced data requirements [16,25,40]

Before delving into this paper, we aim to clarify some typical terminologies related to PIML in BPS, as different studies have used varying definitions to describe similar concepts. To avoid confusion, we briefly introduce three commonly used terms, emphasizing that they all belong

to PIML:

1) **Physics-informed neural networks:** These models are primarily designed to solve PDE problems by embedding governing equations, initial conditions, and boundary conditions into the loss function. This approach incorporates domain-specific knowledge directly into the model training process, helping it better align with physical principles.
2) **Physics-consistent neural networks:** These models ensure physical consistency by incorporating hard structural or parameter constraints. This guarantees that the model complies with physical laws, so any change in the input results in corresponding changes in the output that align with the underlying physics, making them commonly used in dynamic system modeling and control optimization.
3) **Physics-inspired/incorporated/integrated/guided/aware neural networks:** These models follow a similar concept, in which a traditional neural network is "inspired" by physics priors, embedding physical principles into its architecture or design to enhance its alignment with physical realities.

*1.1. Statement of contribution*

Despite its potential, PIML in BPS remains an emerging field, with only a limited number of studies conducted thus far. A recent study [41] provides a structured overview of commonly used PIML methods in building energy modeling, categorizing them into physics-informed inputs, physics-informed loss functions, physics-informed architectural designs, and physics-informed ensemble models. This work serves as a valuable reference for understanding the strengths, limitations, and implementation of various PIML techniques within the context of building energy modeling.

In contrast, our review aims to provide a broader and more critical perspective on PIML across the entire BPS domain. Specifically, we extend the discussion beyond model development to include the definition, motivation, methodology, application areas, current challenges, and future research directions. Our review highlights how PIML can shape the future of BPS workflows and enable next-generation modeling and control strategies. Overall, this study presents a systematic overview of key opportunities, challenges, and a structured roadmap for advancing PIML in BPS. The specific contributions of this study include:

1) **Establishing a clear definition and categorization:** This study gives a clear definition of PIML in BPS and categorizes the commonly used PIML approaches into four types:1) physics-informed data set; 2) physics-informed loss functions; 3) physics-informed model structures, and 4) physics-informed hard constraints. This categorization provides a structured understanding of how physics is integrated into machine learning models.
2) **Reviewing and comparing applications:** This study systematically reviews current applications of PIML in BPS, which include energy prediction and analysis, control and system optimization, IEQ assessment (thermal comfort, IAQ, daylight, acoustics), airflow simulation, and BTM DERs integration. Additionally, it compares PIML with traditional BPS approaches regarding data requirements, model complexity, predictive performance, and computational cost, highlighting its advantages and limitations.
3) **Providing practical resources and guidelines:** To support future researchers, this study summarizes the available datasets, open-source packages, and testbeds for implementing PIML in BPS. Furthermore, it provides practical guidelines for selecting appropriate PIML approaches based on modeling objectives and the required physics knowledge.
4) **Identifying challenges and future directions:** This study thoroughly identifies existing research barriers and outlines potential future directions, establishing a foundation for advancing the development and deployment of PIML in BPS.

A detailed literature review method is provided in **Appendix A**, and the remainder of the paper is organized as follows: Section 2 defines PIML and summarizes commonly used PIML approaches and verification methods. Section 3 reviews major applications of PIML, summarizes available resources, and compares them with traditional BPS methods. Section 4 provides an in-depth discussion, including guidelines for selecting an appropriate PIML approach, identification of current challenges, and proposed future directions. Finally, Section 5 concludes the paper.

**2. Physics-informed machine learning: definition, approaches and verification**

*2.1. Definition of physics-informed machine learning*

PIML is a hybrid modeling framework that integrates fundamental physical laws with advanced machine learning techniques. Unlike traditional data-driven models that rely solely on observed data, PIML embeds physics-based principles and expert knowledge—such as conservation laws, symmetries, and causal relationships—directly into the learning process. This integration is achieved by incorporating governing equations or domain-specific knowledge into model architectures, loss functions, model parameters, or training algorithms. Consequently, PIML enhances the predictive capabilities of machine learning models by improving generalization to unseen scenarios, ensuring physical consistency, and reducing data requirements.

*2.2. Key approaches to incorporate physics into machine learning*

Several methods have been proposed to integrate physics and domain knowledge into machine learning models, focusing on aspects such as physics-informed datasets, model structures, loss functions, and hard model constraints. Kashinath et al. [37]. summarized ten approaches for incorporating physics into machine learning models for weather and climate process prediction. Nghiem et al. [42]. provided a tutorial-style overview of recent advances in PIML for dynamical system modeling and control. Legaard et al. [43]. reviewed various methodologies for constructing dynamical system models using neural networks, while Cai et al. [44]. explored PIML applications in fluid mechanics. Additionally, Karniadakis et al. [25]. offered a general overview of PIML. Readers interested in a broader understanding of how constraints and physics are integrated into machine learning are encouraged to refer to these domain-agnostic PIML review papers (Table 1).

This study, however, focuses specifically on PIML approaches in the context of BPS. These approaches can be broadly categorized into four types: **1) Physics-Informed Data Set, 2) Physics-Informed Loss Functions, 3) Physics-Informed Model Structures, and 4) Physics-Informed Hard Constraints**. The details of each approach are summarized in the following subsections. An overview of PIML approaches and their corresponding applications can be found in Table 2.

*2.2.1. Physics-informed data set*

As shown in Fig. 2A, by its original definition, a physics-informed data set, also referred to as "observational biases" [25], involves how data that inherently reflects physical principles can be used to bias ML models toward predictions that respect those principles, even without being explicitly programmed to do so. In other words, given sufficient training data covering the input domain of a learning task, a machine learning model could potentially learn the underlying physical principles. Such data can be developed through various approaches, including experimental design and dataset development based on prior physics knowledge (e.g., mathematical models), such as combining data from physics-based simulations with measured data or designing

Table 2
PIML applications for BPS are categorized by building type, simulation scope, application, prediction task, modeling method, and results. Abbreviations are used to save space and are listed at the bottom of the table.

| Ref. | Bldg. | Scope | App. | Task | Method | Result |
|---|---|---|---|---|---|---|
| Shao et al. [45] | U | M | Airflow | V, P | L | RMSE<5, 1–2 order faster than CFD |
| Rui et al. [46] | 3D | M | Airflow | V, P | L | MSE 30 % lower than typical PINN |
| Zhang et al. [47] | A | S | Airflow | Infil | S | MAE 43 % lower than gray-box model |
| Mei et al. [48] | U | M | Airflow | C | L | Time: 33.5 h to 2.86 h, NMSE 0.01–0.04 |
| Zhang et al. [49] | 2D | S | Airflow | V, P | L | 34.6 %–53.2 % RMSE reduction |
| Gao et al. [50] | 2D | S | Airflow | V, P | L | MSE 0.3–1.9 %, 3–6 order faster than CFD |
| Hu et al. [51] | 2D | S | Airflow | V | L | $R^2$ 0.67–0.98 with 3, 7 training datasets |
| Rui et al. [52] | 2D | S | Airflow | V, P | L | MSE<0.2 |
| Son et al. [53] | O | S | Airflow | Infil | L | $R^2$ 0.89 |
| Wei and Ooka [54] | 2D | S | Airflow | V, P | L | MAE 55 %–93 % lower than ANN |
| Wei and Ooka [55] | 2D | S | Airflow | V, P | L | RMSE 70 % lower than ANN, 42 % faster |
| Chen et al. [56] | O | S | DER | T | L | MAE 0.25 °C, Peak load ~40 % deduction |
| Xiao and You [57] | R | M | DER | T, H | S, C | Cost saving~39 %, comfort improve~79 % |
| Liang et al. [58] | O | M | DER | T | S, C | 22 % energy saving, SC, SS improve 20,17 % |
| Li et al. [59] | H | \ | DER | T | L | MAE:0.3 °C, $R^2$:0.996 |
| Pandiyan et al. [60]. | H | \ | DER | T | L, S | RMSE reduced 2.57 °C to 2.19 °C |
| Li et al. [61] | H | M | BEM | FDD | I | 95 % accuracy, 2 % higher than DBN |
| Zhang et al. [62] | H | M | BEM | FDD | L | Accuracy 59 % higher than GP |
| Ren et al. [63] | H | M | BEM | FDD | L | Improve detection rate by 27.2 % |
| Michalakopoulos et al. [64] | R | M | BEM | Load | L | NRMSE:0.065, $R^2$:0.87 |
| Ma et al. [40] | E | M | BEM | Load | I | ~90 % CVRMSE lower than physics model |
| Chen et al. [65] | C | \ | BEM | Emission | L | $R^2$:0.96, MAE:12.3 |
| Drgoňa et al. [66] | R | M | Control | T | L, S | Not quantified |
| Bünning et al. [67] | R | S | Control | T | S, C | Energy saving: 26~49 % |
| Wang and Dong [68] | O | S | Control | T, CO2 | S | MAE: 0.4 °C 21 PPM, energy saving:48.9 % |
| Xiao and You [69] | O | M | Control | T, H | S, C | MAE: ~0.3 °C, Energy saving: ~8.9 % |
| Wang et al. [70] | O | S | Control | T, CO2 | S | Accuracy improve 42 % over RC model |
| Wang and Dong [71] | O | M | Control | T, CO2 | L, S, C | 57, 48, 26 % saving by MPC, RL, DPC |
| Nagarathinam et al. [72] | O | M | Control | T | L, S | Prediction error < 1 %, Saving: 12 % |
| Saeed et al. [73] | R | S | Control | T | L | MAE: <1.5 °C, efficiency improvement: 50 % |
| Pavirani et al. [74] | R | S | Control | T | L | 32 % MAE lower than NN, Saving: 9 % |
| Nagarathinam et al. [75] | O | S | Control | T, H | L | MAPE of 0.4 %, Energy saving 16 % |
| Wang et al. [76] | O | M | Control | T | L, S, C | MAPE < 2 %, Energy saving: 33 % |
| Jiang et al. [77] | O | S | Control | T | L, S, C | MAE 0.41 °C, Energy saving: 34 % |
| Drgoňa et al. [78] | R | M | IEQ | T | L, S | MSE of 0.59 K |
| Zhou et al. [79] | O | S | IEQ | Comfort | I | Reduced RMSE by 18.5 % compared to PMV |
| Di Natale et al. [80] | R | S | IEQ | T | S, C | MAE 40 % lower than RC model |
| Gokhale et al. [81] | R | S | IEQ | T | L | MAE: 0.2 °C, Reduced training data:15 % |
| Nguyen et al. [82] | H | M | IEQ | T | L, S | $R^2$ > 0.99 |
| Jaffal [83] | O | S | IEQ | Comfort | I | $R^2$ > 0.9994 |
| Di Natale et al. [84] | R | M | IEQ | T | S, C | MAE: 1.17 °C, MAPE: 4.9 % |
| Yang et al. [85] | C | M | IEQ | T | L, S | RMSE: 0.0012, Reduced data: 60 % |
| Lee et al. [86] | O | M | IEQ | T | L, S | RMSE improvement: 44.7 % over NN models |
| Labib et al. [87] | O | S | IEQ | Light | L | MAE 0.5–0.7, 112 h faster |
| Cho et al. [88] | 2D | S | IEQ | S | L | MSE: 9e-5, 46,000 less data points |
| Karakonstantis et al. [89] | O | S | IEQ | S | L | RMSE: 2 dB |
| Olivieri et al. [90] | 3D | S | IEQ | S | L | NMSE: 6.71 dB |
| Chen et al. [91] | R | S | IEQ | T | L | CVRMSE of 4.31 % and RMSE of 0.94 °C |
| Chen et al. [92] | D | S | IEQ | T | S, C | Reduced 79.2 % MAE |
| Jiang and Dong [16] | R, O | S,M | IEQ,BEMControl, DER | T, H Load | L, S, C | MAE: 0.43 °C, $R^2$: 0.91 for load prediction, 43 % energy saving, 55 % load shifting |

Abbreviations:.

'Bldg.'/Building Type: 'R' (Residential Building), 'O' (Office Building), 'A' (Apartment), 'U' (University), 'H' (HVAC System), 'C' (Commercial Building), 'D' (Data Center), 2D/3D (2D/3D Simulation Geometry).

'Scope'/Simulation Scope: 'U' (Urban-scale), 'M' (Multi-zone), 'S' (Single-zone).

'App.'/Application: 'Airflow' (Airflow Simulation), 'DER' (Behind-the-Meter DERs Integration), 'BEM' (Energy Prediction and Analysis), 'Control' (Controls and System Optimization), 'IEQ' (IEQ Assessment).

'Task'/Prediction Task: 'V' (Velocity), 'P' (Pressure), 'Infil' (Infiltration), 'C' (Concentration), 'T' (Temperature), 'H' (Humidity), 'FDD' (Fault Detection and Diagnostics), 'Comfort' (Thermal Comfort),'Light' (Lighting), 'Load' (Load Prediction), 'S' (Sound Field).

'Method/PIML Method: 'I' (Physics-Informed Data Set), 'L' (Physics-Informed Loss Functions), 'S (Physics-Informed Model Structures), 'C' (Physics-Informed Hard Constraints).

'Result': 'SC' (Self-Consumption), 'SS' (Self-Sufficiency). Self-Consumption describes the portion of generated renewable energy (e.g., from PV panels) that is directly used by the building rather than exported to the grid or curtailed. Self-Sufficiency represents the percentage of the building's total energy demand that is met by on-site renewable generation. We have now added these definitions to the manuscript for clarity.

well-structured experimental datasets based on domain expertise.

This method is also referred to as "knowledge discovery" [93,94] or "physics-informed learning", meaning that machine learning can help scientists discover new insights by learning from large amounts of observational data. However, this does not mean that any ML model trained using such data, such as that generated by EnergyPlus, or experimental data should be considered a PIML model. In the context of BPS, current PIML efforts are primarily focused on "knowledge embedding" [93], which refers to incorporating domain knowledge into data-driven models to enhance their accuracy, robustness, efficiency, and consistency. In this context, simply learning from data without integrating physics-based priors is no different from a traditional data-driven model and should not be classified as PIML. Only when data from physics-based models or physics priors directly contribute to joint predictions can it be considered PIML.

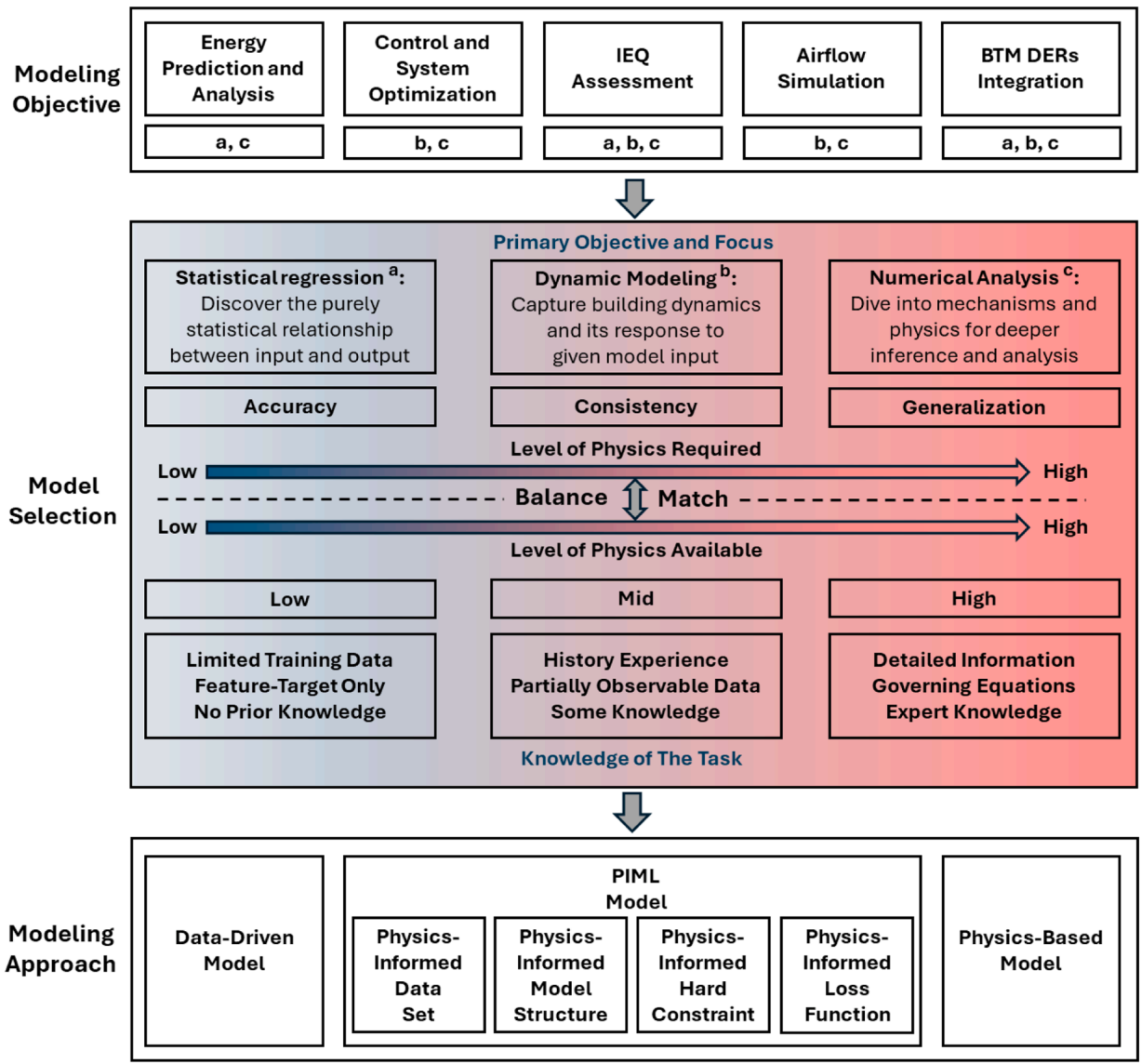

**Fig. 1.** A general PIML pipeline for BPS: from understanding the modeling objective to identifying required physical principles, available physics knowledge, and selecting the appropriate modeling approach.

For example, Ma et al. [40]. proposed an ensemble learning framework that divides the prediction task into a physics-driven component, using EnergyPlus for HVAC load calculations, and a data-driven component, where an LSTM model predicts residuals related to occupancy-driven variations and noise based on observed data. Jaffal et al. [83]. developed a PIML thermal comfort meta-model for non-air-conditioned buildings, where features were determined by quasi-steady-state heat transfer principles and fitted using simulation data in a polynomial form. Zhou et al. [79]. proposed a hybrid model to predict personalized thermal comfort. A regression model was initially used to estimate non-observable personal parameters such as metabolic rate, skin temperature, and saturated vapor pressure. These outputs were then integrated with a physics-based thermal comfort model to predict the thermal comfort level. Vaghefi et al. [95]. integrated a physics-based model with a data-driven time-series model to forecast building energy usage, capturing physical priors through a set of zonal energy balance equations with estimated parameters.

Despite its success in improving model performance, ensemble learning relies heavily on physics-based models. Unlike other approaches that integrate physics knowledge directly into machine learning models and train them in parallel, ensemble learning follows a sequential training process. A physics-based model must be developed and calibrated, and this dependence conflicts with the primary goal of PIML, which is to enhance model scalability. Moreover, the data-driven models used here are still constrained by the previously mentioned limitations, as no physical priors have been incorporated directly into the data-driven models.

Another example of physics-informed model inputs can be found in Li et al.'s work [61], where a physics-guided Bayesian diagnostic network was developed based on expert knowledge as priors, and a genetic algorithm was applied to further optimize its structure using operational data.

*2.2.2. Physics-informed loss functions*

As shown in Fig. 2B, the physics-informed loss function-also known as physics-based regularizations—involves integrating physics-informed constraints as penalty functions into the training process. During training, the machine learning (ML) model not only minimizes a standard loss, such as mean squared error (MSE) or cross-entropy loss, to reduce the discrepancy between ground truth and predictions but also incorporates a customized physics-informed loss function to enforce adherence to underlying physical principles. This approach can be viewed as multitask learning [96], where the physics-informed loss imposes 'soft' constraints, and the relative weight of the physics-based penalty compared to the standard loss is treated as an adjustable hyperparameter. By tuning these soft penalty constraints, the model balances fitting observed data and respecting the underlying physical laws.

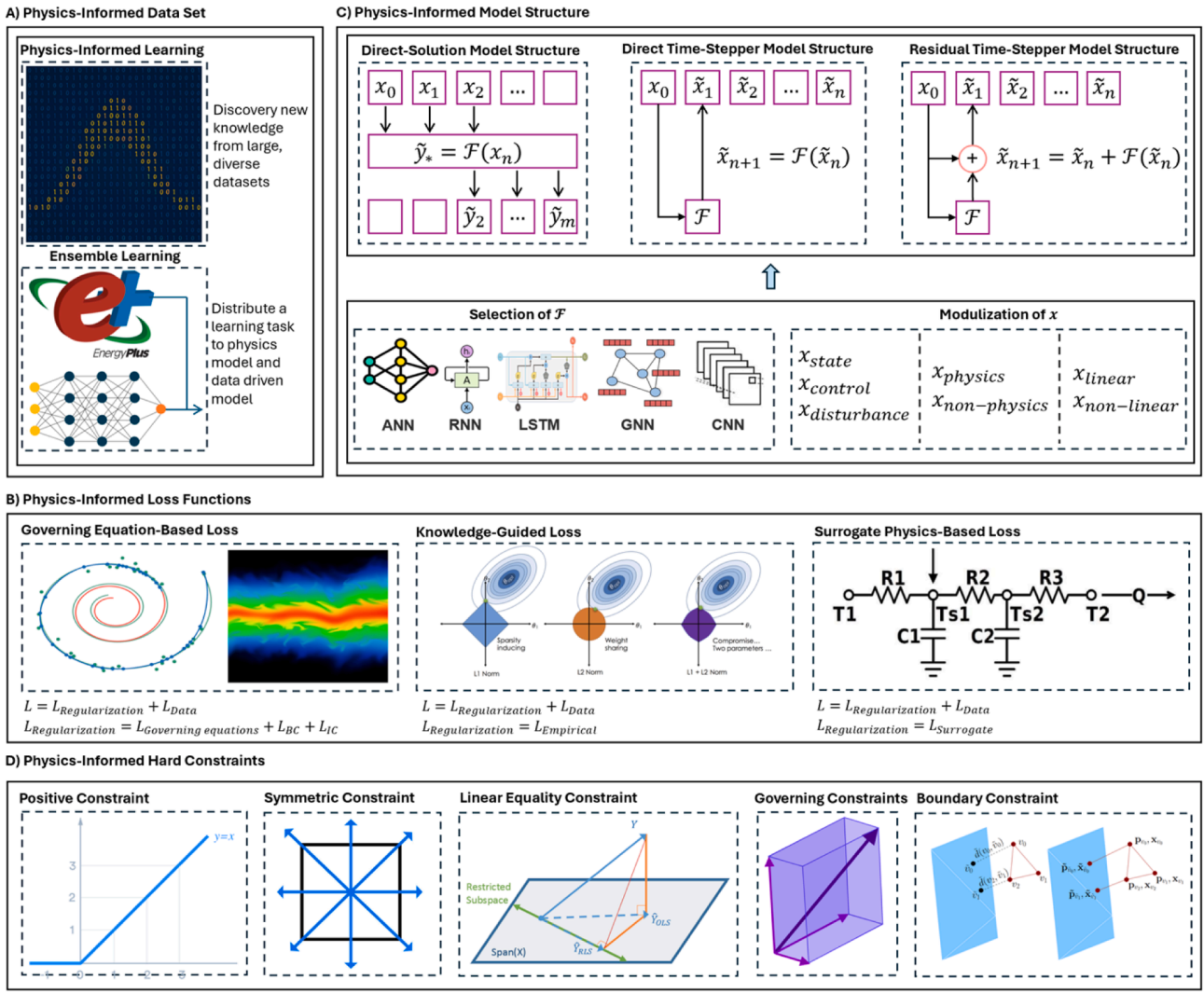

Fig. 2. Overview of physics-informed model learning approaches.

In general, physics-informed loss functions can be categorized into three types:

**1) Governing equation-based loss**

This type of loss enforces physical constraints derived directly from governing equations (e.g., differential equations) that describe system dynamics. A widely used method within this framework is the physics-informed neural network (PINN), initially introduced to solve forward and inverse problems involving nonlinear partial differential equations (PDEs). By incorporating losses from boundary conditions, initial conditions, and governing equations, PINNs have demonstrated broad applicability in domains such as fluid dynamics, biology, quantum mechanics, and reaction-diffusion systems [36]. Detailed information about PINNs can be found in studies [44,96,97].

**2) Knowledge-guided loss**

In real-world applications, governing equations or boundary conditions are often unavailable or difficult to define; therefore, domain knowledge is derived from expert experience rather than explicit equations. For example, negative power consumption in building energy modeling is penalized to ensure physically realistic predictions.

**3) Surrogate physics-based loss**

Instead of directly encoding physics into the loss function, a surrogate physical model is developed to generate additional data by providing pseudo-labels for unobservable states and guiding the learning process. For example, a reduced resistor-capacitor (RC) model can be used to guide the training of a building dynamic model.

Below is a summary of their applications:

**1) Governing equation-based loss**

Shao et al. [45]. proposed a physics-informed graph neural network for rapid urban wind field prediction, where the loss function comprises two components: a regular loss measuring the distance between predictions and observed data, and a physics-based term constrained by the Reynolds-averaged Navier–Stokes (RANS) equations. Rui et al. [46]. predicted the 3D flow field around a building using a physics-informed neural network, where the loss function was adjusted according to the residuals of the governing equations, boundary conditions, and training data, respectively, based on the RANS equations. Similarly, Mei et al. [48]. designed a physics-informed neural network with a PDE-based loss function for pollutant concentration prediction. Labib [98] integrates the radiative transfer equation and the diffusion equation with MSE loss to predict building daylighting performance. Cho et al. [88]. incorporated acoustic wave equations into their neural network, allowing it to learn time-varying acoustic waves, including reflections, in grids of arbitrary geometry, with the acoustic source being a sinusoidal point source. Similarly, Karakonstantis et al. [89]. integrated the inhomogeneous wave equation with neural network to construct the early part of the room impulse response while completing sound field characterization in the time domain. Li et al. [59]. embedded the PDE residual loss into a neural network to govern heat transfer in a borehole thermal energy storage system. Chen et al. [65]. integrates the inference of physical laws into the loss function to estimate building carbon emissions. Pandiyan et al. [60]. customized the loss function by embedding differential equations to respect the heat transfer dynamics in an electric water heater.

**2) Knowledge-guided loss**

Drgoňa et al. [78]. introduced an inequality constraint to penalize the unrealistic model response (such as a 1 K change in ambient temperature causing a 2 K change in indoor temperature within a single time step). They also designed an eigenvalue penalty as a U-value emulator and a state difference penalty to produce smoother state trajectories that stabilize the dynamic system. Similarly, Nguyen et al. [99]. introduce a smoothness penalty term and boundary violation loss to ensure the smoothness and boundedness of the proposed model. Wang et al. [68]. add a regularization loss on the model parameters to learn the weight decay behavior that can compel the model to focus on closer inputs. Ren et al. [63]. developed a physics-informed loss function that included an autoencoder-induced loss by comparing squared residuals between the inputs and outputs with detection thresholds, as well as a thermodynamic law-induced loss based on mass balance, energy balance, and redundancy for fault detection. Jiang et al. [16,77]. incorporated a regularization loss with their encoder and the decoder's first step output to provide a stable initialization, along with another regularization loss to penalize incorrect model responses, such as temperature increases with additional cooling. Yang et al. [85]. developed a graph neural network-based building thermal dynamics model, where the rate of temperature change in each room is assumed to be proportional to the area. A corresponding loss function was included as a penalty term to reduce variations during model training. Lee et al. [86]. integrated physics-informed flow and energy loss at critical junctures involving complex interactions among various HVAC components to learn real-world behaviors and adhere to the physical laws. Zhang et al. [47]. customized a decision tree model for building infiltration prediction. The proposed model incorporated Coblenz and Achenbach's infiltration equations into the training loss to optimize the tree structure, such as the number of "leaves" and "branches." Zhang et al. [62]. presented a physics-informed Gaussian process model for HVAC system performance prognosis. Physics priors were embedded by incorporating analytical equations as the mean function of the Gaussian process, while the kernel functions were designed to capture time-varying degradation effects.

**3) Surrogate physics-based loss**

Gokhale et al. [81]. integrated an RC model with a neural network. Building on this framework, Nagarathinam et al. [72,75]., Gokhale et al. [81]., Chen et al. [56, 91]., and Pavirani et al. [74]. adopted similar approaches. In these studies, an RC building thermal model was first developed using prior knowledge of the building. In some cases, the model required pre-calibration, while in others, it was calibrated jointly during neural network training. The model was then used to predict room temperature and lumped thermal mass temperature. These predictions served as training labels to compute the MSE loss between the latent state and the unobservable variables in the neural networks. This regularization loss was then integrated into the deep learning framework, guiding the network to learn system dynamics and latent representations simultaneously.

While all three approaches integrate physical insights into model training, they differ in how they incorporate physics, whether through explicit equations, expert experience, or surrogate models. Generally, governing equation-based loss provides the most rigorous physical consistency. In contrast, knowledge-guided loss and surrogate physics-based loss offer greater flexibility in real-world scenarios with incomplete or uncertain physical information.

*2.2.3. Physics-informed model structures*

Physics-informed model structures offer a powerful way to integrate physical principles into ML frameworks, providing deeper insights into traditionally opaque "black-box" models. These structures enhance both the interpretability and performance of data-driven models. In Fig. 2C, we summarize three commonly used computational schemas. For most BPS applications, these schemas fall into two categories: **direct-solution models** and **time-stepper models** [43]. Additionally, we identify five widely used NN models—ANN, RNN, LSTM, GNN, and CNN—each suited to specific modeling tasks. For instance, ANN offers high computational efficiency, RNN and LSTM handle temporal dependencies, while GNN and CNN capture spatial relationships, making them suitable for multi-zone tasks. These models are employed to calculate latent representations $\mathscr{F}$, as shown in Fig. 2C. To enhance model performance, input features are often modularized based on domain knowledge. Common strategies include splitting data into state variables, control inputs, and disturbances to mimic state-space models; separating physics-based and non-physics-based components for independent processing; or dividing inputs into linear and nonlinear components to build specialized modules. Detailed information on these methods is provided in the sections below.

1) **Direct-solution models**, learn hidden representations $\mathscr{F}(x)$, where $x \in \mathbb{R}^{n \times d}$, $\mathscr{F} : \mathbb{R}^{n \times d} \to \mathbb{R}^{n \times m}$ from training datasets. These models predict target variables based on input features through a learned input-output mapping. For example, in an estimation problem, the input and output can share the same time index. However, in a forecasting problem, the output may have a different time index depending on the forecasting horizon. Regardless of the specific task, once trained, direct-solution models can be applied at any desired time step. Wang et al. [68,71]. developed a partially connected and input convex neural network using a mask matrix. To mimic a state space formulation, a shared weight mechanism was applied along the model diagonal and its parallels. Lee et al. [86]. proposed an encoder-decoder structured diffusion convolutional recurrent neural network for building dynamic modeling, where prior knowledge of the floor plan was embedded in a graph structure to capture heat transfers between adjacent zones. The major difference between a direct-solution model and a time-stepper model is that the direct-solution model can directly predict the value of the next state without going through a recursive loop at each time step, which is more computationally efficient, while a time-stepper model is more physically meaningful.
2) **Time-stepper models**, can be considered as numerical solvers and are commonly used for dynamic system modeling. Based on current or past state measurements; these models estimate future system states over a defined time horizon. This model can be further categorized into the following two types:

**2.1) Direct time-stepper model**. In this model, the next state is predicted by the network directly from the current state. For example, Drgoňa et al. [78]. developed a state-space-informed model structure in which the state, input, and disturbance dynamics were represented by a separate neural network individually. In this way, the model can capture the highly nonlinear thermal dynamics and impose structural assumptions and constraints due to the decoupling structure. Xiao et al. [69,100]. developed a two-layer NN combining an LSTM cell and an RNN cell. Due to the Hadamard product computation in LSTM, ensuring physics consistency is challenging. Therefore, the proposed model uses the LSTM cell to handle inputs that do not require physics consistency, such as the occupant number, solar radiances, and time information while using the RNN to process physically consistent variables, such as temperature, humidity, and HVAC power. Bünning et al. [67]. developed a physics-informed autoregressive–moving-average model with exogenous inputs and compared its control performance with two other machine learning models. Separate modules were designed to estimate solar gains, adjacent heat transfer, ambient temperature, and actuator gains, based on linearized physics equations.

**2.2) Residual time-stepper model**. In this mode, a network is trained to predict the step-wise change (i.e., a derivative-like residual quantity) in the system state over a time step. This residual is then added to the previous state to compute the next state. For example, Di Natale et al. [80,84]. proposed a module-based model structure that incorporates a linear energy accumulator to represent energy changes driven by external temperature and HVAC power. Additionally, a black-box module was developed to capture the effects of nonlinear terms. This work was later extended to multizone studies by incorporating an adjacent heat transfer module. Similarly, Jiang et al. [16,77]. developed a modularized neural network to estimate each heat transfer term of a building dynamic system based on energy balance equations. While an encoder is designed to extract historical information, a current cell measures the current time step, and a decoder predicts system responses based on future system inputs and disturbances. Furthermore, a graph neural network is employed to combine multiple single-zone models for multi-zone modeling tasks. One notable difference between these two models lies in the coupling of the disturbance module with the state variable. The former decoupled model is more control-friendly and compatible with convex linear programming optimization, but might sacrifice some accuracy, particularly in buildings with heavy thermal mass.

*2.2.4. Physics-informed hard constraints*

As shown in Fig. 2D, unlike physics-informed loss functions, which act as soft constraints by guiding the model toward satisfying physical principles without enforcing strict adherence, physics-informed hard constraints ensure that the model's outputs strictly comply with underlying physical mechanisms. Several methods have been proposed to enforce such hard constraints. For example, Lu et al. [101] applied an adaptive Lagrangian method that dynamically adjusts the penalty coefficient to improve constraint satisfaction. Richter-Powell et al. [102]. introduced a divergence-free neural network that always satisfies the continuity equation. Négiar et al. [103] learned a family of functions mapping PDE parameters to PDE solutions and identified an optimal linear combination from this function family by solving a PDE-constrained optimization problem, which naturally enforces the physical constraints. Saad et al. [104] developed a Boundary-Enforcing Operator Network that satisfies Dirichlet and Neumann boundary conditions by structurally modifying the operator kernel. Mayr et al. [105]. introduced Boundary Graph Neural Networks, which dynamically modify graph structures to obey boundary conditions. Chalapathi et al. [106]. applied a Mixture-of-Experts mechanism to impose constraints over smaller, decomposed domains, with each subdomain solved by an expert network through differentiable optimization. Chen et al. [107] proposed KKT-hPINN, which strictly enforces hard linear equality constraints using projection layers derived from Karush-Kuhn-Tucker conditions.

In building thermal dynamic modeling, the concept of physical consistency was first introduced by Di Natale et al. [80]., who defined a BEM model as physically consistent with respect to a given input when any change in that input leads to a change of the output that aligns with the underlying physical laws. For example, a decrease in cooling power at previous time steps should lead to an increase in indoor temperature at the present timestep —and vice versa for heating. This constraint can be mathematically defined as shown in Equation(1):

$$\frac{\partial y_t}{\partial u_k^{HVAC}} > 0, \ \forall 0 \leq k < t \qquad (1)$$

Where $y$ represents space air temperature, $u^{HVAC}$ means HVAC power, $t$ and $k$ are time step. More detailed information can be found from [80].

Di Natale et al. [80,84]. developed three linear modules to model heat transfer through the environment, adjacent zones, and the HVAC system, respectively. Each module was constructed using non-negative parameters to ensure compliance with physics consistency constraints. Xiao et al. [69,100]. introduced a RNN module designed to capture physically consistent inputs, such as HVAC power, outdoor temperature, and humidity. To meet these requirements, they enforced the positivity of inputs and initial weights to comply with Equation(1). They also explained their preference for RNN over LSTM models, due to the challenges posed by the Hadamard product employed in LSTM structures. Similarly, Wang et al. [71]. and Jiang et al. [16]. incorporated non-negative constraints on the weights of their models to preserve the physics consistency of HVAC systems, the environment, and neighboring zones.

It is worth noting that enforcing positive gain is only one example of applying hard constraints. In the BPS domain, there are broader opportunities —for example, equality constraints for energy and mass balance, symmetry constraints in adjacent heat transfer, and boundary condition constraints in CFD simulations. All of these can be treated as hard constraints and appropriately integrated into PIML models.

*2.2.5. Summary of physics-informed machine learning approaches*

At the end of this section, we summarize the different PIML approaches, their required prior knowledge, suggested applications, and associated pros and cons in Table 3.

*2.3. Model verification*

Studies have shown that deep neural networks involve more hyperparameters than shallow ones, increasing the probability of overfitting [108] and leading to poor generalization performance, particularly when applied outside the training domain. For example, a small change in input can cause significantly biased results [109]. This instability poses a major barrier to the real-world implementation of deep learning models in BPS, especially for control-oriented applications, which heavily rely on accurately predicting system responses to given control inputs.

As discussed earlier in previous studies [16,69,80,84,110,77], while classical LSTM models can achieve reasonably accurate predictions (as shown by the gray line in Fig. 3B), they often fail to capture the underlying physics—such as the effects of heating and cooling on temperature. A consistent neural network should respond appropriately to changes in input. For example, in Fig. 3A, when the HVAC power is intentionally set to different levels, the room temperature should change accordingly. However, as shown in Fig. 3B, a regular LSTM model without physics-consistency constraints fails to follow this trend. The reason behind this is that, even when the training data is sufficient in quantity, its coverage is often not broad enough—especially in the building domain, where HVAC systems typically operate within a

Table 3

Summary of Physics-Informed Machine Learning approaches, required knowledge, suggested applications, and their pros and cons.

| Approaches | Categorization | Required Knowledge | Suggested Applications | Pros | Cons |
| --- | --- | --- | --- | --- | --- |
| Physics-Informed Data Set | Physics Informed Leaning | Large, diverse datasets to discover new physical insights | Exploring new mechanisms in BPS | Inspire discovery of unknown physics | Requires large volumes of high-quality data |
| | Ensemble Learning | Pre-defined physics model | Reducing uncertainty error not well captured by physics models | Improve model performance | Requires additional calibration of the physics model, physics is not directly embedded into the data driven model |
| Physics-Informed Loss Functions | Governing Equation-Based Loss | Governing equations with boundary/initial conditions of the system | Problems governed by well-defined physics equations (e.g., heat transfer, CFD) | Improved data efficiency, fine-grained output | Challenging to balance multiple loss terms, struggles with complex boundary conditions (e.g., nonhomogeneous problem) |
| | Knowledge-Guided Loss | Expert domain knowledge or empirical rules | Lacks well-defined governing equations, but empirical knowledge is available | Flexible to implement and effectively regulates model behavior | Embedded knowledge may be biased |
| | Surrogate Physics-Based Loss | Well defined physics surrogate model | Dynamic modeling where surrogate model canimprove structural consistency | Guide model learning in latent space | Requires additional calibration of physics model, limited scalability |
| Physics-Informed Model Structures | Direct-solution models | Clear understanding of the modeling objective and appropriate architecture selection (e.g., ANN, RNN, CNN, GNN) | Tasks with specific temporal/spatial physical requirements (e.g., dynamic modeling) | Computationally efficient | Less adaptive; struggle with long simulations or complex systems |
| | Time-stepper models | | | Highly flexible to state and boundary condition changes; physically interpretable | Accumulates errors, stability concern |
| Physics-Informed Hard Constraints | Positive constraints | Governing equations with boundary/initial conditions of the system | Tasks with strict physical requirements (e.g., energy control optimization) | Strictly enforces physical rules | Limited solution space could degrade model performance |
| | Symmetric constraints Linear equality constraints Governing constraints Boundary constraints | | | | |

narrow comfort range. This limitation prevents the model from accurately capturing the true relationship between system inputs and outputs, especially under unseen conditions, and may lead to controller failures.

To mitigate overfitting, common training strategies such as cross-validation and early stopping are often applied. However, in the BPS domain, these methods only focus on accuracy evaluation while overlooking consistency. Therefore, this section summarizes the available metrics and methods beyond accuracy for evaluating control-oriented, data-driven building dynamic models as shown in Table 3, which is a critical step before deploying them in real-world scenarios.

The commonly used data-driven performance metrics include mean absolute error (MAE), mean absolute percentage error (MAPE), mean squared error (MSE), and root mean squared error (RMSE), as shown in Table 2. These metrics primarily focus on accuracy and are widely applied in regression tasks, such as building load forecasting and temperature prediction. However, when evaluating a model's physical consistency, the focus must shift toward whether the model responds correctly to given control inputs. Here, we summarize several approaches in Table 4:

**1) Model sanity check** [16,69,71,80,84,68,110,77]

To verify whether a model's output is physically consistent, a basic approach is called sanity check. Fig. 4 illustrates an example of evaluating a model's consistency with respect to its control input—HVAC power. In this case, the objective is to determine if the model responds correctly to varying HVAC inputs. For example, the sign of the system's gain, as well as reasonable response speed and amplitude. To test this, the HVAC power is intentionally reset (e.g., to its maximum, minimum capacity) starting from the third timestep(or any other), while all other features remain unchanged.

Qualitatively, we can manually evaluate whether the response aligns with our prior knowledge. For instance, under maximum cooling input, the room temperature is expected to decrease and eventually converge to a threshold at a certain decay rate. However, such information is often challenging to observe directly in real-world settings, making the evaluation highly dependent on expert knowledge. Alternatively, quantitative evaluation can be conducted using the temperature response violation (TRV) metric, as defined in study [110], TRV is calculated by Equation(2) (3):

$$TRV^+ = sum(min(T_{up} - T_{pred}), 0) \qquad (2)$$

$$TRV^- = sum(min(T_{pred} - T_{down}), 0) \qquad (3)$$

Here, $T_{pred}$ represents the predicted results based on the original input, while $T_{up}$ and $T_{down}$ are the predicted results under the modified (sanity check) inputs. This metric quantifies the degree of consistency violation in a given data-driven model. And a physically consistent model should exhibit zero violations.

However, the TRV method only verifies the sign of the system's gain but does not capture the magnitude or speed of the response. For example, with increased cooling input, one model might predict a 1 K temperature decrease within 15 min, while another predicts the same decrease over 60 min. Although both models may yield a TRV of zero, the model closer to the real system is preferable. Therefore, another metric —maximum mean discrepancy (MMD) [110] —can be used to quantify how consistent the data-driven model is compared to the real dynamic system. MMD measures the distance between two sample distributions by calculating the maximum difference in sample averages over a kernel function. The key idea is to perform multiple sanity checks to generate a sample distribution of the system's responses to various inputs and then compare this distribution to the measured response distribution. The smaller the MMD value, the closer the model aligns with the real system.

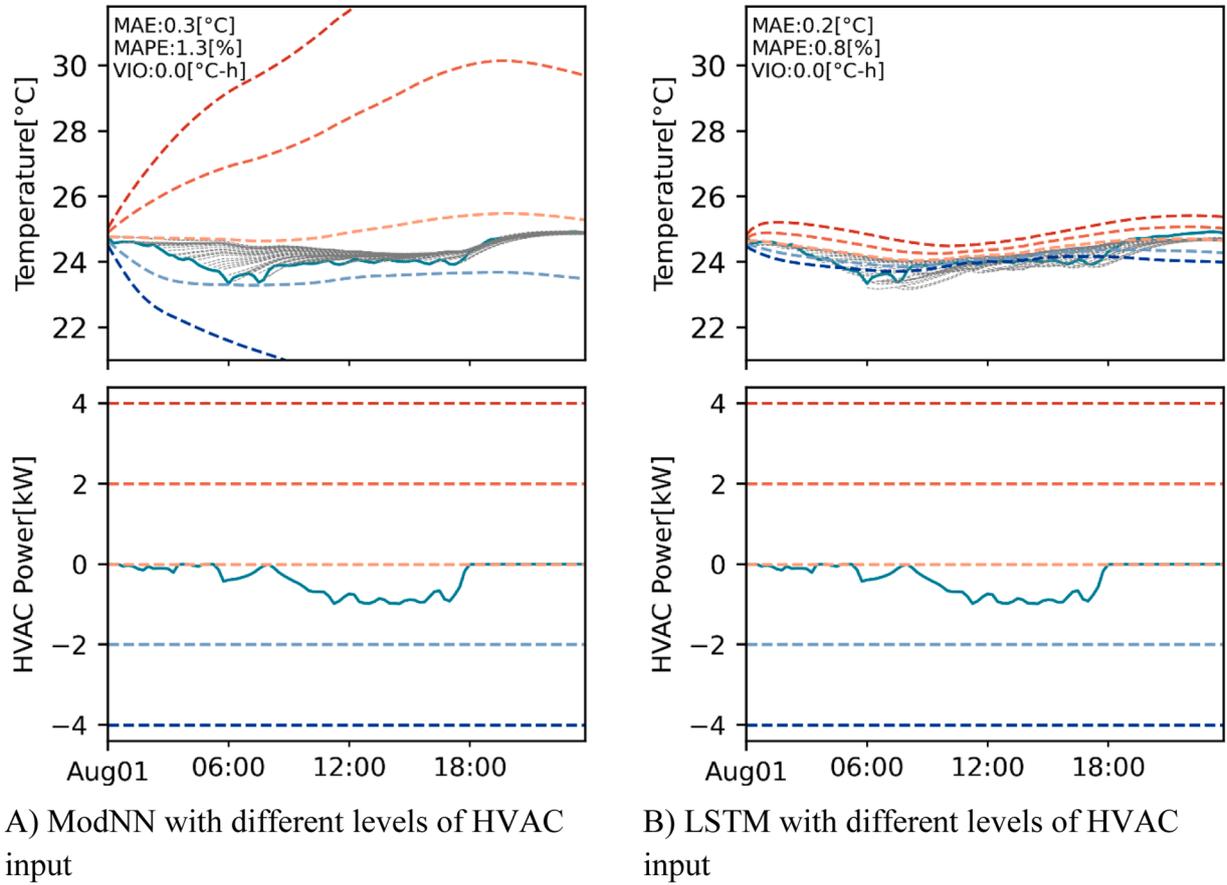

Fig. 3. Example [77] of model responses with different levels of HVAC input.

**Table 4**
Commonly used method to evaluate physics consistency for control-oriented model.

| Applications | Objective | Method | Reference |
| --- | --- | --- | --- |
| Control and System Optimization | Sign Consistency of System Gain | Sanity Check | Ref [16,69,71, 80,84,68,110] |
| | Magnitude Consistency of System Gain | Gradient Evaluation | Ref [16,71,84] |
| | Feasibility and stability | Eigenvalues | Ref [69,78,100] |

**2) Gradient evaluation** [16,71,84,77]

Gradient evaluation is another method to assess the physical consistency of a PIML model. The Jacobian matrix of the model outputs with respect to input features can be easily computed through automatic differentiation (AD) in existing packages like PyTorch and TensorFlow. Depending on the applications, these gradients can be either qualitatively or quantitatively evaluated. For indoor temperature prediction, non-negative gradients are usually expected. For instance, as outdoor temperature or HVAC load increases, the indoor temperature should also increase. In multi-step indoor temperature predictions, the gradient

| Feature \ Timestep | 1 | 2 | 3 | 4 | ... | 96 |
| --- | --- | --- | --- | --- | --- | --- |
| $u^{HVAC}$ | $P_1^{HVAC}$ | $P_2^{HVAC}$ | $P_{min/max}^{HVAC}$ | $P_{min/max}^{HVAC}$ | | $P_{min/max}^{HVAC}$ |
| $w^{Amb}$ | $T_1^{Amb}$ | $T_2^{Amb}$ | $T_3^{Amb}$ | $T_4^{Amb}$ | | $T_{96}^{Amb}$ |
| $w^{Adj}$ | $T_1^{Adj}$ | $T_2^{Adj}$ | $T_3^{Adj}$ | $T_4^{Adj}$ | | $T_{96}^{Adj}$ |
| $w^{Sol}$ | $P_1^{Sol}$ | $P_2^{Sol}$ | $P_3^{Sol}$ | $P_4^{Sol}$ | | $P_{96}^{Sol}$ |

**Model Inputs**

Fig. 4. Example of a model sanity check for control input.

should increase due to thermal lag—especially in heavyweight buildings—and diminish gradually from closer time steps to farther ones, reflecting the temporal dependency. In fluid-dynamics-related studies, gradients evaluation can help verify adherence to governing equations, boundary conditions, and initial conditions—ensuring the model respects physical principles.

**3) Eigenvalue evaluation** [69,78,100]

Eigenvalues are used to verify the feasibility and stability of a PIML model. Where feasibility means there exists a feasible solution and stability refers to the requirement for bounded outputs and states with bounded inputs, which is crucial for system safety. According to Hautus Lemma that a matrix pair ($A$,$B$) is stabilizable if and only if it satisfies Eq. (4):

$$rank[\lambda I - A, B] = n, \forall \lambda \in \{\lambda | \lambda \in \Lambda, Re(\lambda) \geq 0\} \quad (4)$$

Where ($A$,$B$) can be derived from the model parameters and $\Lambda$ is the set of all eigenvalues of $A$.

## 3. Physics-informed machine learning in building performance simulation: applications, resources, and comparisons

### 3.1. Current applications of physics-informed machine learning in building performance simulation

This section systematically analyzes the state-of-the-art applications of PIML by categorizing them according to fundamental building physics laws. As shown in Fig. 5, among the reviewed studies, **PIML for IEQ assessment** emerges as the most extensively studied category. These applications include temperature modeling (T), humidity modeling (H), $CO_2$ modeling, lighting simulation (L), and sound simulation (S), and thermal comfort (TC). The second most common application area is **control and system optimization**, with most studies focusing on model predictive control (MPC), differentiable predictive control (DPC), reinforcement learning (RL), and rule-based control (RB) at the building level. Another major area of application is **airflow simulation**, where PIML is employed to simulate fluid dynamics, including velocity (V), pressure (P), and concentration (C) distributions. Six studies have extended this work to **BTM DERs integration**, covering not only HVAC control but also water tanks (WT), batteries and PV systems, water heaters (WH), and borehole thermal energy storage systems. There are also a few studies focused on **building energy prediction and analysis**. In this context, prediction refers to predicting future energy usage (e.g., load forecasting, energy consumption estimation), while analysis involves interpreting historical or real-time data to identify energy usage patterns, evaluate system performance (e.g., fault detection and diagnosis, FDD), or conduct primary energy assessments such as emissions calculations.

#### 3.1.1. IEQ assessment

Recent advances in PIML have demonstrated significant potential in building IEQ assessment, including temperature, humidity, $CO_2$ concentration, lighting, and acoustic predictions. And most of them are focused on temperature prediction as shown in Fig. 5.

Drgoňa et al. [78]. developed the first PIML model for 30-day multizone temperature predictions, achieving an MSE of 0.59 K with only 10 days of training data, compared to 1.07 K for the best linear model based on the Perron-Frobenius theorem informed model constraint. Di Natale et al. [80]. outperformed an RC baseline with an MAE of 0.88 °C for 72-hour predictions (RC: 1.48 °C) and extended their framework to multi-zone tasks [84] achieving an MAE of 1.17 °C (RC: 1.79 °C; LSTM: 1.27 °C). Nagarathinam et al. [75]. improved LSTM results with a MAPE of 0.4 % for single zone air temperature and 0.2 % for wall temperature prediction (LSTM: 5 % and 3 %, respectively). Then further extend their work [72] for multizone tasks, which reduced errors to ~1 % (PINN baseline: 9 %). Wang and Dong [68,71] achieved an average MAE of 0.44 °C for temperature and 21 ppm for $CO_2$ predictions, later extending their model to multizone applications with MAPEs of 1.14 %–1.96 %. Xiao and You [69] developed a multizone temperature model with relative error below 5 %, slightly higher than LSTM due to model constraints. Jiang and Dong [16,77] attained MAEs of 0.43 °C (RC: 0.94 °C) for single-zone and 0.27 °C–0.39 °C for multizone tasks, with humidity predictions reached MAEs of 0.0002–0.0005 kg water/kg air. Lee et al. [86]. achieved an RMSE of 0.78 K for multi-zone temperature prediction, outperforming LSTM (1.51 K) and GRU (1.48 K). Other single-zone studies, including Bünning et al. [67]. Gokhale et al. [81]. and Chen et al. [56,91]. achieved similar performance.

PIML has also shown promise in simulating and reconstructing acoustic fields within buildings. Cho [88] applied PIML to simulate time-varying acoustic wave behavior in enclosed spaces. By incorporating boundary conditions and solving wave equations, the model captured detailed sound wave reflections. The approach demonstrated computational efficiency compared to traditional solvers, but its focus on 2D geometries and simplified sinusoidal sources highlighted a need for expansion to more complex scenarios. Similarly, Karakonstantis et al. [89]. explored room impulse response (RIR) reconstruction in reverberant environments, using PIMLs to estimate sound pressure and velocity fields. Their study focused on a specific experimental room and achieved a 10 dB reduction in RMSE compared to state-of-the-art methods —even with sparse observations. Despite these successes, challenges remained in extending the approach to diffuse reverberant fields and diverse room configurations. Expanding to volumetric sound field modeling, Olivieri et al. [90]. reconstructed 3D acoustic fields in recording studios. By embedding wave equations into the training process, the PIML accurately predicted sound pressure across time (NMSE =

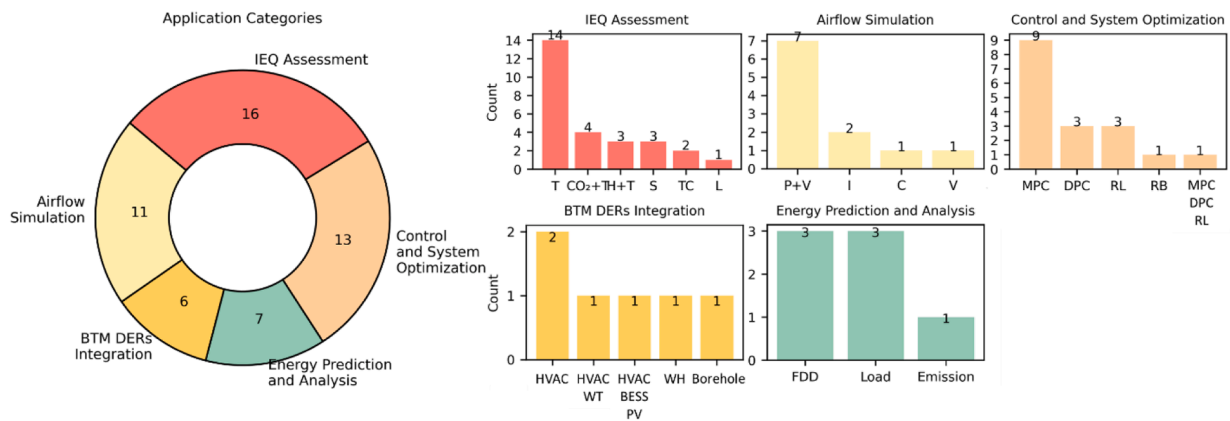

**Fig. 5.** Current applications of PIML in BPS.

−20.71 dB) and space while maintaining computational efficiency. The method achieved improved accuracy over traditional solvers and reduced dependence on dense microphone arrays, demonstrating robustness in controlled environments. However, its application to dynamic or highly reverberant spaces remains an area for future research.

In a separate application, Labib [87] developed the first PINN for predicting daylight autonomy in buildings. The Radiative Transfer Equation (RTE)-based model achieved an MAE of 0.70, while the diffusion-based model demonstrated improved accuracy with an MAE of 0.55.

*3.1.2. Control and system optimization*

One of the primary objectives of building indoor environment modeling is to enable advanced control and optimization. Building on the models discussed above in Section 3.1, several studies have extended these approaches to support control optimization, as illustrated by the red line in Fig. 6.

Drgoňa et al. [66]. demonstrated the utility and scalability of their modeling approach for learning neural control policies for real-time HVAC system optimization. They implemented the approach with a DPC controller to minimize energy use while maintaining the desired thermal comfort levels. The proposed method effectively constrained control actions (e.g., mass flows) for each output, achieving near-optimal performance trajectories. However, the model did not account for physical consistency with thermal dynamics constraints.

To overcome this limitation, Di Natale et al. [80] incorporated physics-based hard constraints into their model. Their approach ensured that the gradients of temperature predictions at the end of the prediction horizon, with respect to power inputs and external temperatures observed along the horizon, remained non-negative. Building on this model, Liang et al. [58]. applied an MPC controller based on a tropical office building. The PCNN-enabled MPC achieved a 27 % reduction in peak demand and a 22 % reduction in energy consumption compared to the baseline control. Additionally, the approach enhanced the building's self-sufficiency and PV self-consumption by 17 % and 20 %, respectively.

Similarly, Xiao and You's model incorporated physics-based hard constraints, the proposed model was implemented with MPC for energy optimization [69] and one-day-ahead energy management [57]. In the first case, it reduced energy consumption by 5.8 % to 8.9 % and improved thermal comfort by 55 % to 64 % compared to conventional controllers. In the second case, the proposed method achieved 29.5 % to 39.7 % energy cost savings and improved thermal comfort by 65.4 % to 79.3 % compared to other methods, respectively. However, the model's high nonlinearity and complexity required energy optimization to be solved by particle swarm optimization (PSO), which requires a high computation cost.

To enhance computational efficiency, Wang and Dong [68] reformulated the energy optimization problem into a convex form, significantly reducing computational costs. The proposed model, implemented with MPC, achieved average reductions of 36 % in space cooling loads and 72.8 % in airside coil energy consumption. The model was further refined for physical consistency and validated through a long-term experimental study [71] using MPC, DPC, and RL, realizing energy savings of 56.98 %, 48.39 %, and 30.6 %, respectively. Building on this framework, the model was adapted for multi-zone dynamic prediction in an office building and integrated with RL [76]. This adaptation resulted in a 33 % reduction in radiant panel energy use compared to the baseline, while maintaining zone temperatures within the specified comfort ranges. At the same time, Pavirani et al. [74]. explored the benefits of a PIML simulator within a Monte Carlo Tree Search (MCTS) to enhance state forecasting and controller performance in residential heating systems. The study demonstrated the PIML's capacity to reduce energy costs by 4 %. Gokhale et al.'s model [81] integrated with an MPC controller and achieved a 32 % reduction in state forecasting error, a 4 % reduction in energy consumption, and a 7 % improvement in thermal comfort.

Jiang and Dong [16] developed an encoder-decoder structure to effectively handle historical states and ensure stable initial conditions. Based on a modularized model structure grounded in energy balance principles, the model guaranteed physics consistency and was implemented with a DPC controller. The proposed model achieved a 43.3 % reduction in energy consumption, reduced peak loads by 55.3 %–95.1 %, and lowered energy costs by 28.1 %–79 %. This model was further improved [77] and integrated with RL, demonstrating a 34 % energy-saving potential in a three-month experimental study.

*3.1.3. Energy prediction and analysis*

This section focuses on FDD and load prediction. For FDD, Li et al. [61]. developed a hybrid Diagnostic Bayesian Network (DBN) for fault detection and diagnosis (FDD) in air handling units (AHUs). By combining operational data with expert knowledge, the model achieved a detection rate (DET) and diagnosis rate (DIA) of 95.37 %, with an error rate (ERR) of 0.73 %. Local causal graphs provided interpretable visualizations, aiding expert evaluations. While validated experimentally, further research is needed to assess scalability and real-world applicability in dynamic environments. Zhang et al. [62]. proposed a physics-guided Gaussian Process (PGGP) model for HVAC system performance prognosis. Compared to standard Gaussian Process (GP), Support Vector Machine (SVM), and Recurrent Neural Network (RNN) models, the PGGP improved accuracy by 11.4 %, 31.7 %, and 5.0 %, respectively. Additionally, the PGGP demonstrated superior data efficiency, outperforming standard GP, SVM, and RNN models by up to 59.9 %, 75.5 %, and 68.4 %, respectively. Ren et al. [63]. developed a deep learning model integrated with thermodynamic laws for high-dimensional sensor fault detection. The proposed method improved the fault detection rate by 27.2 % and significantly reduced the false alarm rate by 77.4 %.

For load prediction task, Michalakopoulos et al. [64] developed a physics-informed DNN that achieved an $R^2$ of 0.87±0.01 and an RMSE of 102.69±7.82. Ma et al. [40]. outperformed traditional physics-based models, achieving 40–90 % improvements in both MAE and CV-RMSE. Similarly, Jiang and Dong's [16] model demonstrated strong generalization capabilities with an overall $R^2$ ranging from 0.79 to 0.94 and MAE between 0.11 kW and 0.73 kW under various conditions. Additionally, this model is also the first study to predict post-retrofit energy consumption based on operational data.

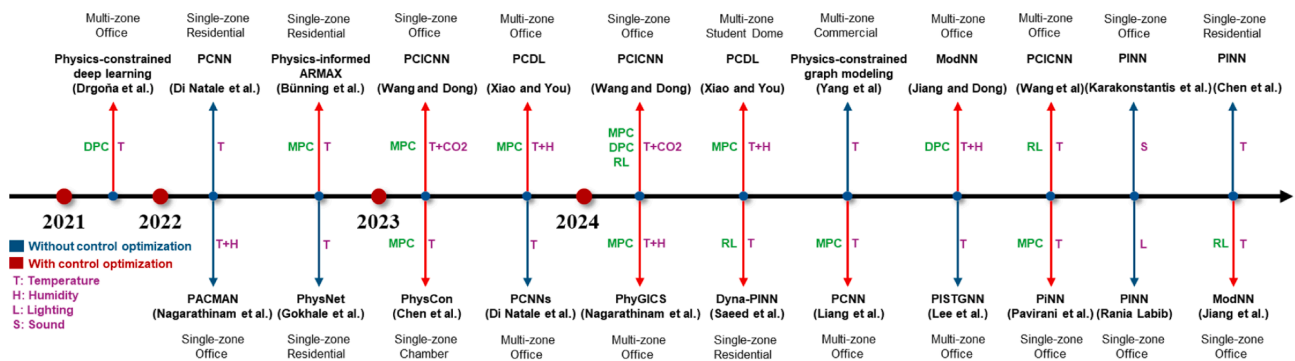

**Fig. 6.** Development timeline of representative PIML models in IEQ assessment.

*3.1.4. Airflow simulation*

Another widely studied application of PIML within the broader scope of BPS is airflow simulation. Reconstructing detailed indoor airflow fields from limited measurement data remains a critical challenge. Purely data-driven methods, such as traditional neural networks, often lack physical interpretability and struggle under data scarcity, while traditional CFD simulations are time-consuming and heavily rely on accurate boundary conditions. PIML offers a promising alternative by combining the advantages of traditional physics-based simulations with the adaptability of data-driven methods, addressing limitations inherent in both. Moreover, PIML approaches frequently incorporate experimental data and numerical simulations, facilitating the development of robust and generalizable models for various purposes, such as thermal condition evaluation, ventilation design, and pollutant dispersion assessment. In this section, we categorize the PIML applications in airflow simulation into two scales: zone scale (space air distribution) and city/urban scale (wind field simulation) as follows:

**1) Zone Scale**

Wei et al. [111]. introduced a PINN to reconstruct airflow fields in small, two-dimensional indoor environments. By leveraging sparse measurement data and incorporating governing equations into the training process, the model reduced error rates by up to 70 % compared to traditional artificial neural networks, decreased computing time by 42 %, and enhanced predictive capabilities for pressure fields. However, the simplified assumptions of steady-state, isothermal conditions and two-dimensional geometry limit its applicability to real-world indoor airflow scenarios, which often involve more complex dynamics.

In another recent study, Son et al. [112]. developed a PINN model to estimate ventilation and infiltration rates in residential and small commercial spaces. By incorporating long-term observational data, the model captured temporal variations in operations and meteorological conditions. The results—validated using real-world measurements—highlighted key influences on air change rates, such as window opening behavior, wind direction, and speed. Their findings also highlight the potential of PINNs to enhance the understanding of airflow phenomena alongside their predictive capabilities. Despite these advantages, the study relied on uniform indoor $CO_2$ concentration assumptions and did not incorporate outdoor $CO_2$ data, which could affect model accuracy. Future research should explore more complex geometries, integrate outdoor measurements, and evaluate the use of multiple indoor sensors.

**2) City/Urban Scale**

Urban wind field prediction is another critical area for PIML, particularly in addressing challenges such as urban heat island effects and pedestrian comfort. Rui et al. [52]. applied a dynamic prioritization PINN (dpPINN) to model three-dimensional wind fields around building clusters in urban-scale environments. The dpPINN leveraged sparse near-wall measurement data, offering a practical approach for real-world applications. By introducing a self-adaptive loss strategy to balance multi-objective optimization during training, the model demonstrated superior accuracy and computational efficiency, achieving results 1–2 orders of magnitude faster than traditional CFD simulations. These advancements make dpPINN a promising tool for reconstructing airflow in urban contexts.

Similarly, Shao et al. [45]. developed a physics-informed graph neural network, PIGNN—CFD, to predict urban wind fields over complex layouts. Their study integrated CFD-generated training data with sparse real-world measurements, enabling the model to accurately reconstruct airflow in large-scale urban areas. The PIGNN—CFD model incorporates the RANS equations to ensure physical consistency while maintaining high performance even with limited data. By reducing computational costs by 1–2 orders of magnitude compared to traditional CFD, the model offers a scalable solution for urban-scale airflow modeling. The study highlights the potential of PINNs to balance data-driven and physics-informed approaches effectively. Future work should focus on exploring transfer learning capabilities and further reducing training and computational costs to broaden the applicability of PINNs in urban environments.

*3.1.5. BTM DERs integration*

BTM DERs integration presents a promising solution for enhancing grid flexibility. The application of PIML-based building dynamic models enables optimization at a large scale for building energy systems, facilitating centralized coordination and control across diverse buildings. Such centralized strategies are important for achieving efficient resource allocation and dynamic grid interactions.

Several studies have demonstrated the potential of leveraging advanced modeling frameworks for this purpose. For example, Chen [56] and Jiang [16] focused on HVAC control to improve grid flexibility. Using learned building dynamics, Chen developed a rule-based controller by adjusting temperature setpoints in response to grid demands while Jiang developed a DPC controller for demand response by precooling strategy. Liang et al. [58]. extended their framework to include BESS-PV systems and similarly, Xiao and You [100] incorporated additional water tank systems into their framework that reduced energy consumption ranging from 29.5 % to 39.7 % and improve thermal comfort level up to 79 %. Both approaches demonstrated significant load-shifting potential, contributing to improved grid responsiveness and energy management. However, current studies primarily use PIML to represent building dynamic models and can be considered as a subclass of control and system optimization as discussed in Section 3.2. Future research could further expand PIML frameworks to explicitly include components for renewable generation, energy storage [59,60], and HVAC components such as chillers and heat pumps, enhancing its scalability and applicability.

*3.2. Available resources to support research and development in physics-informed machine learning*

This section summarizes available resources in the BPS domain that can support future research and development of PIML. We focus on three key aspects: datasets, packages and testbeds.

*3.2.1. Datasets*

As PIML is still an emerging field, its performance under diverse conditions—such as varying climates, weather patterns, building types, and HVAC systems—remains an open question. To better understand the strengths and limitations of PIML compared to traditional ML models, we summarized several open-source datasets that can be used for model training and benchmarking in Table 5. For example, the ÉPFL team [80] released a dataset containing over three years of data collected from an apartment in the NEST testbed in Dübendorf, Switzerland. Miller et al. [113]. published an open data set from 507 non-residential buildings that includes hourly whole building electrical meter data for one year. New York State Energy Research and Development Authority (NYSERDA) has provided an open dataset covering over 300 buildings in New York State [114]. This dataset includes data from building automation systems, connected devices, utility meters, equipment sub-meters, and IoT sensors. The ASHRAE Great Energy Predictor III dataset [115] offers data from over 1000 buildings over a three-year period, encompassing weather data as well as electricity, chilled water, steam, and hot water meter readings. LBNL released a 3-year timeseries dataset of >300 sensors and meters for a medium-sized office building in Berkeley, California [116]. Ecobee [117] has contributed an open-access dataset collected from smart thermostats installed in 1000 single-family homes across four states, covering the entire year of 2017. This dataset includes information on system and equipment operations, indoor and outdoor environmental conditions, occupant behavior, and building system assets. LBNL also developed a synthetic dataset of medium-sized office building operation performance [118], which includes HVAC, lighting, miscellaneous electric loads (MELs) system operating conditions, occupant counts, environmental parameters, end-use and whole-building energy consumptions at 10-minute intervals, covering 30 years' historical weather data in three typical climates of Miami, San Francisco, and Chicago.

**Table 5**
Available datasets for PIML in the BPS.

| Dataset | No. of BLDGs | BLDG Location | BLDG Type | Time Span | Time Resolution | Measurement |
|---|---|---|---|---|---|---|
| NEST Testbed [80] | 1 | Dübendorf, Switzerland | Combined office & residential units | 3 years | 15 min | • Indoor temperature<br>• Weather conditions<br>• Electricity |
| The Building Data Genome [113] | 507 | USA, EU, Singapore, Australia | Non-residential, Most education | 1 year | Hourly | • Electricity |
| RTEM [114] | ~300 | New York State, USA | Non-residential | 5 years, Varies for different buildings | 5 min, 15 min, Hourly | • System operation<br>• Electricity<br>• Indoor environment |
| ASHRAE - Great Energy Predictor III [115] | 1000 | Not mentioned | Education, Office, Public, Residential | 3 years | Hourly | • Electricity<br>• Water-side energy<br>• Weather conditions |
| LBNL [116] | 1 | Berkeley, California, USA | Office | 3 years | Minutely, 5-minute, 10-minute, 15-minute | • Indoor temperature<br>• Electricity<br>• Indoor occupancy |
| Ecobee DYD [117] | 1000 | California, Texas, New York, and Illinois, USA | Single family home | 1 year | 5-minute | • System operation<br>• Indoor Environment<br>• Occupant<br>• Weather condition |
| Synthetic building operation [118] | Not apply | Miami, San Francisco, Chicago, USA | Office | Not apply | 10-minute | • Energy consumption<br>• Power demand<br>• Indoor environment<br>• Occupancy<br>• System operation |

### 3.2.2. Open-source packages

For instance, Raissi et al. [119]., Drgona et al. [120]., Zubov et al. [121]., Lu et al. [122]., Ehsan et al. [123]. have developed PINN solvers for addressing forward and inverse problems. These tools are primarily based on physics-informed neural networks with loss functions incorporating partial differential equations. They can be particularly useful in the building physics domain, including applications such as envelope heat transfer, airflow modeling, lighting simulation and more. However, some of these solvers are limited to well-defined physical problems and may struggle with complex or highly nonlinear systems. Another tool called AutoKE [124], which can automatically embed equations of arbitrary forms into computational graphs and has a broader range of applications. In building energy modeling domain, Di Natale et al. [80,

84]., Gokhale et al. [81]., Chen et al. [56]., Jiang et al. [16] have open-sourced PIML models for building dynamic modeling. These models are based on customized building architectures or constrained model parameters, enabling fast prediction of space air temperature and humidity using building operation data. They also support advanced building energy control applications. To promote adoption and reproducibility, many of these tools provide detailed documentation and example scripts in their respective repositories. Interested readers are encouraged to visit the GitHub pages, where step-by-step examples can be found to guide the application to their own problems.

### 3.2.3. Testbeds

To evaluate and benchmark proposed PIML models, several open-source testbeds are available. For instance, the Building Controls Virtual Test Bed (BCVTB) enables co-simulation by linking various simulation tools, including EnergyPlus, Modelica, Radiance, and MATLAB/Simulink, for co-simulation. It can also connect these programs to Building Automation Systems and databases for model-based operation [125]. Building Operation Testing Framework (BOPTEST) [126] is another powerful testbed which contains a runtime environment to start a building emulator, set up tests, control the virtual systems, access data, and report KPIs. Additionally, CityLearn [127] provides a Multi-Agent Reinforcement Learning (RL) for building energy coordination and demand response in cities. COBS [128] is another open-source, modular co-simulation platform for developing and comparing building control algorithms, which integrates various simulators and agent models with EnergyPlus and supports fine-grained and occupant-centric control of building subsystems [129]. These testbeds can be utilized to benchmark control strategies developed using PIML techniques.

## 3.3. Comparison between physics-informed machine learning and traditional BPS approach

In this section, we compare PIML with traditional physics-based and data-driven BPS approach from the following aspects:

### 3.3.1. Data requirement

Physics-based models typically require comprehensive metadata about the modeled system. For example, building energy modeling requires detailed information on space layout, orientation, and envelope properties, while CFD simulations require geometric details. Such requirements are often difficult to obtain in practical applications due to limited access to accurate data. Data-driven models, on the other hand, rely heavily on the quantity and quality of the dataset. The dataset should cover a wide range of potential scenarios and include essential features and labels that capture key aspects of the problem. This ensures the model can learn patterns and make reliable inferences based on the data provided. PIML significantly reduces the amount of training data required compared to traditional data-driven models. By leveraging prior knowledge from physical laws and equations, PIML effectively guides the learning process and constrains the solution space, improving efficiency and reducing the dependence on extensive datasets. However, PIML requires additional physics-related information, which is often difficult to obtain and might be imperfect or biased. For example, common assumptions such as well-mixed air, homogeneous material properties, or clean and continuous boundary conditions may not hold in real-world scenarios. These imperfect physics data can lead to degraded model performance.

### 3.3.2. Modeling effort

Physics-based models often require the greatest effort to develop. In the BPS domain, creating a detailed geometry model is typically the first step, which takes substantial time. Depending on the specific modeling task, additional upfront information is also required. For instance, building energy simulations involve developing a layer-by-layer building envelope model and defining the schedule for each system. Similarly, CFD models require careful meshing and proper boundary and initial conditions to ensure convergence. In real-world applications, due to the unavailability of many physical parameters, these models often require repeated and case-by-case calibration to meet testing thresholds. Data-driven models typically involve processes such as data cleaning, feature engineering, model development, training, validation, and testing. They provide a straightforward solution for mapping data inputs to outputs, learning system behaviors without extensive manual input from modelers. However, due to poor generalization ability, traditional data-driven models often require additional rounds of learning when applied to new datasets or tasks. In contrast, PIML benefits from incorporating prior knowledge of physical laws, which significantly enhances the model's generalization ability. However, the integration of physics priors often increases model complexity due to multiple loss components and potentially conflicting constraints, making the tuning process more challenging than in traditional data-driven methods.

### 3.3.3. Model performance

The performance of a physics-based model is highly dependent on the modeler's domain knowledge and the quality of input data. While physics-based models can generalize well when properly calibrated, their performance may still be affected by modeling assumptions, the level of detail in the data, and simplifications. Biased observations—such as oversimplified boundary conditions—can introduce epistemic uncertainty due to incomplete or inaccurate system representations.

Data-driven models, in contrast, learn from data. They often achieve high accuracy under normal operating conditions but may suffer from limited generalization ability and lack physical consistency, leading to incorrect predictions under unseen scenarios. In these models, predictive uncertainty largely depends on the quality and coverage of the training data.

PIML improves model performance by embedding physical knowledge into data-driven models. These physics-guided constraints narrow the solution space, enhancing the model's ability to generalize to out-of-sample conditions and ensuring responses that are physically consistent. This effectively reduces epistemic uncertainty. For example, Jiang et al. [16]. evaluated their PIML model under a disruptive power outage event and demonstrated that the model could still correctly capture the temperature response trend despite the deviation from normal operating conditions. Similarly, Li et al. [61]. incorporated expert knowledge into a Diagnostic Bayesian Network, significantly improving diagnostic accuracy and reducing uncertainty.

However, aleatory uncertainty—from the stochastic nature (e.g., occupant behavior)—cannot be fully addressed through physics embedding. While PIML improves consistency and generalization, it does not inherently reduce aleatory uncertainty. To address this, probabilistic methods such as Bayesian neural networks could be explored. It is also worth noting that, due to the added physical constraints, PIML models may have slightly lower accuracy than purely data-driven models on standard test datasets, as the constrained solution space. Particularly, recent studies have shown that PINNs have not outperformed traditional numerical solvers such as the finite element method when applied to PDEs [130], especially in nonhomogeneous problems. PINNs often struggle with abrupt changes in material properties at interfaces, such as those encountered in multi-layer heat transfer problems [131].

### 3.3.4. Computation cost

Physics-based models do not require training but rely on solving forward physics equations. Their computation time depends on model complexity, such as geometric details, meshing resolution or simulation duration, and can grow exponentially as complexity increases. Data-driven models, while requiring significant time for training, benefit from rapid inference speeds. Training time depends on factors such as model structure, dataset size, batch size, and the number of training epochs. But with the development of GPU technology, the training time has significantly reduced. And during inference, data-driven models are highly efficient. PIML models have similar inference speeds to purely data-driven models but often require longer training times. This is due to their more complex structures, which integrate both data-driven components and physical knowledge, increasing the computational demands during training.

## 4. Discussion

### 4.1. A general guideline for selecting appropriate physics-informed machine learning model

PIML offers a promising approach for advancing BPS. However, integrating PIML into every BPS application is neither feasible nor effective due to the inherent complexity of the model and the higher computational costs compared to purely data-driven methods. To develop suitable models, researchers must adopt a structured approach that balances modeling objectives with available resources. This section provides a three-step guideline for selecting appropriate PIML models based on the nature of BPS applications, the required level of physics, and the level of available physics knowledge.

**Step 1: Understand the modeling objectives and the required level of physics**

As shown in Fig. 1, BPS applications can be categorized into three types based on their modeling objectives and associated physics needs:

1) **Statistical regression:** Focused on identifying the purely statistical input-output relationship, this category prioritizes result accuracy without addressing the internal dynamics of the system. Such as load prediction, occupancy prediction and weather forecasting. Such objective requires the lowest level of physics knowledge.
2) **Dynamic modeling:** Aims to capture the time-dependent behavior of a system, requiring both accuracy and consistency. These tasks require a moderate level of physics knowledge, since these models must respect underlying physical laws and ensure stability over time, such as control-oriented building dynamic modeling and surrogate modeling.
3) **Numerical analysis:** Targets detailed insight into physical processes and system mechanisms. These tasks require high physics fidelity and often prioritize inference generalizability over predictive performance. For example, thermal transfer analysis and CFD simulations.

Step 2: Identify the available physics knowledge.

Next, researchers should assess what prior knowledge and data are available for the modeling task. For example, clear knowledge of the modeling task, e.g., information about the building (geometry, envelope, occupancy schedule, HVAC system). Availability of boundary conditions, governing equations, and expert knowledge. Dataset including relevant features, targets, length and diversity (coverage).

Step 3: Match Appropriate PIML Methods to Knowledge Levels

Based on the required and available physics knowledge levels, the next step is to select an appropriate PIML model that balances these factors. To support this selection process, we clarify the differences and principles of primary PIML methods as follows:

1) **Governing equation-based loss:** This approach incorporates the highest level of physics knowledge, as it explicitly derives loss functions from governing equations that describe a physical mechanism. It requires a well-defined problem formulation, including known boundary conditions and equations. Due to its high fidelity, it is best suited for fine-grained simulations such as CFD, lighting field modeling, acoustics field modeling, and heat transfer analysis.
2) **Knowledge-guided loss:** When governing equations are unavailable or observations are insufficient, an alternative way is to embed physics priors empirically. However, the effectiveness of this approach is not always guaranteed, as human bias may affect the proposed loss function. The effectiveness of this method heavily depends on expert knowledge. Despite its limitations, knowledge-guided loss can improve training efficiency, enhance accuracy, and increase generalizability, making it suitable for tasks with some prior knowledge available.
3) **Model hard constraints:** This method enforces predefined rules within the model, ensuring strict satisfaction with certain physical principles. It is particularly useful for tasks requiring reliable model responses, such as control optimization or tasks requiring high-fidelity predictions such as CFD simulations. However, while hard constraints guarantee rule adherence, they may not always align with real-world scenarios due to human bias, similar to knowledge-guided loss methods. Additionally, since hard constraints directly influence model behavior, they may sometimes degrade performance. Therefore, it is essential to evaluate the effectiveness of proposed physics priors before integrating them into data-driven models. Interested readers may refer to the latest studies [132] that introduce systematic approaches for this evaluation.
4) **Surrogate physics-based loss and physics-informed datasets:** These approaches provide insights into unobservable states and are commonly used in thermal dynamic modeling and FDD. However, developing surrogate models requires additional modeling effort, while generating physics-informed synthetic datasets involves extra simulation rounds, which may reduce scalability, making these approaches more task-specific.

In real-world applications, prior physical knowledge is often imperfect or biased. As shown in a recent study [77], incorporating such rules does not always improve model accuracy or consistency and may even degrade performance. Therefore, it is essential to evaluate whether the incorporated physics rules actually enhances model performance.

Key questions include: How can we determine if a given physics prior is useful? How should we select, update, or replace these priors? How can we combine different rules to further improve model performance? To answer these questions, the third step focuses on evaluating the value of incorporated physics priors [132], including their dependence, synergism, and substitution effects. Such evaluations provide practical guidance on selecting appropriate physics constraints for different tasks, ultimately leading to more robust and reliable physics-informed machine learning models.

By following this guideline, future researchers can effectively match PIML techniques to specific BPS applications.

*4.2. Current research barriers and challenges of physics-informed machine learning*

*4.2.1. Challenges to tradeoffs between physics and data-driven application*

Although several studies have demonstrated that integrating physics prior knowledge can significantly improve model performance, it is crucial to determine the appropriate level of physics integration. Some applications are adequately addressed with either the physics-based models or data-driven models alone; there is no need for PIML models. Knowing when or what use cases can benefit from PIML models rely on the researcher's experience and domain knowledge. Here, we summarize three key drawbacks of an overwhelmed "physics-informed" machine learning model.

**1) Model development**

Incorporating physics knowledge into machine learning models increases the challenges of model development. During the architecture design stage, integrating prior physics knowledge requires a deep understanding of building physics and domain expertise, adding significant complexity compared to a purely data-driven "black-box" model. Furthermore, during the training stage, physics-informed machine learning often involves multiple regularization losses, transforming the process into a multi-task learning problem. This requires careful tuning of the weights assigned to each regularization loss to ensure both performance and stability. Additionally, the physical quantities in BPS can sometimes differ by orders of magnitude regarding their typical values (e.g., indoor temperature is usually within 10 to 40 °C, while heating and cooling loads can exceed 10,000 W). How to scale different data features to stabilize the PINN training, while maintaining the model's physical meaning requires dedicated efforts such as reformulating the governing physical equations, converting units, and tuning weights for different terms in the physical regularizations.

**2) Model Accuracy**

Integrating physics knowledge into model design does not always guarantee improved predictive accuracy. While physics-informed models, such as PINNs, benefit from architectures and regularization constraints that narrow the solution space, these constraints often introduce a trade-off between model expressiveness and physical interpretability. Specialized architectures grounded in simplified governing equations may limit the model's capacity to capture complex, nonlinear relationships inherent in real-world data. This limitation arises because the physical assumptions and regularizations inherently restrict the degrees of freedom in parameterization, compared to the flexibility of purely data-driven, black-box models. While this trade-off ensures that the results align with established physical principles, it may not fully leverage the predictive power of the data, especially when the physical equations used are oversimplified or approximate. Previous studies [69,110,16] have highlighted cases where accuracy was compromised due to these constraints.

### 3) Model Implementation

Building dynamics are inherently nonlinear, and incorporating physics knowledge can make models more representative of real systems. However, this also increases model complexity, leading to higher computational costs. For real-time optimal control problems, complex models may not solve efficiently within the short optimization windows required. Additionally, complex models pose significant challenges for optimization. For instance, Xiao et al. [69]. and Jiang et al. [16] developed models that integrated complex deep learning structures, which could not be solved directly, and they turn to use Particle Swarm Optimization (PSO) or Dynamic Programming Control (DPC) methods. In contrast, Wang et al. [68] and Di Natale et al. [80]. simplified their models—sacrificing some predictive accuracy—to reduce complexity, which allowed their problems to be reformulated as convex optimization problems that are more control-friendly.

*4.2.2. Challenges to integrate physics priors more efficiently*

An open question in the development of PIML is how to integrate physics priors efficiently, in terms of minimizing data requirement, reducing computational costs (both for training and utilization), and designing more targeted model structures and priors. As summarized in Section 2, there are four commonly used physics integration methods in terms of physics-informed data set, physics-informed loss functions, physics-informed model structures, and physics-informed hard constraints. Below, we discuss the limitations of each method:

### 1) Physics-informed data set

In general, a physics-informed data set can be derived through experimental design or generated using simulation tools to inherently adhere to physical laws. However, there is a lack of standardized benchmarks for data collection, which will be discussed in the next Section 4.3.6 on benchmarks. Another approach is to obtain data from simulation tools. As discussed earlier in Section 2.2.1, this method relies on physics-based models, which are often not scalable. The key challenge lies in seamlessly integrating the outputs from physics-based models with PIML frameworks to achieve parallel, scalable, and efficient workflows.

### 2) Physics-Informed Loss Functions

This approach incorporates physics priors through regularization terms in the loss function. Each category presents distinct challenges:

- Governing equation-based losses require precise mathematical formulations of governing equations, along with their boundary and initial conditions, which are often complex or unavailable in real-world applications.
- Knowledge-guided losses depend heavily on the understanding of the system and available auxiliary information (e.g., wall temperature, heat flux, supply air temperature bounds, or open-loop datasets) that can provide learning guideline for underlying dynamics.
- Surrogate physics-based losses, such as those informed by RC models, typically require prior development and calibration. A poorly calibrated RC model fails to guide NN learning, while a well-calibrated one may not need additional NN integration. Calibrating an RC model and then integrating it with an NN contradicts the core objective of PIML. The key challenge is to integrate RC model calibration and NN training in parallel to enhance structural consistency without compromising scalability.

### 3) Physics-informed model structure

How to select, design and assemble physics-informed neurons, layers, or blocks that encode or enforce specific physical properties requires expertise in both data science and building domain knowledge, which presents significant challenges. For instance, graph neural networks and convolutional layers can be utilized to capture spatial relationships within buildings, such as heat transfer through the environment or between adjacent zones. Meanwhile, recurrent neural networks are well-suited for capturing temporal relationships, making them ideal for modeling dynamic systems often represented as time-series data and with temporal dependencies. The key challenge lies in assembling these components appropriately. Each module learns a high-dimensional representation in latent spaces based on customized architectural designs, which are often opaque in black-box models and difficult to verify. Systematically identifying the most suitable neural network components for each physical process and ensuring their proper integration to produce accurate, physically consistent predictions remains an open research challenge.

### 4) Physics-informed hard constraints

Physics-informed hard constraints strictly enforce underlying physical laws but reduce model flexibility and can lead to ill-conditioned training especially for imperfect data with noise. Moreover, enforcing such constraints in complex or large-scale systems significantly increases computational cost. In general, embedding complex physical laws directly into neural networks—especially for general or high-dimensional problems, remains challenging and requires further exploration.

*4.2.3. Challenges to evaluate model performance*

### 1) Evaluation of the physics properties of a physics-informed machine learning model

Most current studies focus on accuracy-oriented performance evaluation metrics, such as MAE and MSE, leaving a significant gap in assessing the model's alignment with physical properties. Questions like "Does the model's response make sense?" or "Is it consistent with the system's behavior?" are often overlooked. A model that predicts accurately but fails to capture the correct response dynamics can lead to significant control failures, as it does not adequately learn the true underlying system dynamics. To address this, we need a physics-based performance metric to evaluate models. Such a metric could also guide the training process. While model accuracy typically improves with more training epochs, however, the consistency with physical principles might decline after a certain threshold, leading to overfitting. A physics-informed model verification framework could help select appropriate models and mitigate overfitting by ensuring both accuracy and consistency with physical properties.

### 2) Model verification

Model verification remains a substantial challenge for physics-informed machine learning models for BPS. As Section 2.3 illustrated, most existing studies relied on historical measurements for model development and off-line testing. While historical data provides a foundation for assessing model performance under typical operational conditions, they often fail to cover the full range of real-world scenarios, particularly extreme or atypical events. Validating models with real-time field measurements remains difficult and is seldom achieved, limiting the ability to ensure that models generalize well beyond the conditions represented in the training data. Furthermore, conducting experiments to evaluate model performance under a broader range of conditions is inherently challenging in real-world settings where operational parameters are tightly controlled to maintain safety and comfort. For instance, in built environments, temperatures are regulated within narrow bounds to ensure occupant comfort, making it difficult to observe and verify model behavior under more extreme conditions like heat waves or cold snaps.

*4.3. Future directions of physics-informed machine learning*

*4.3.1. Enhanced integration of physics and data-driven approaches*

One promising direction is to refine the balance between physics-based and data-driven modeling. Current PIML frameworks often rely on predefined physics priors, which may not fully capture the complexities of real-world scenarios. Future research could focus on adaptive methodologies that dynamically learn the appropriate degree of integration based on the available data and physical constraints. This would enable models to adapt to varying levels of data availability and uncertainty, ensuring robust and scalable solutions across diverse applications.

*4.3.2. Closing the loop between knowledge embedding and discovery*

Future research should focus on establishing a bidirectional feedback loop between knowledge embedding and knowledge discovery to advance PIML. While knowledge embedding [93] integrates domain knowledge into machine learning models to enhance accuracy, robustness, and consistency, knowledge discovery, through methods like symbolic regression and sparse regression, explores governing equations and new scientific insights from observational data. Rather than treating these as separate processes, future studies can explore how discovered knowledge can iteratively refine embedded physics priors, creating an adaptive learning framework that continuously improves both model fidelity and scientific understanding. This closed-loop integration could lead to more interpretable, generalizable, and physically consistent machine learning models, driving innovation in BPS and beyond.

*4.3.3. Advanced neural architectures*

Designing novel neural network architectures that better incorporate domain-specific knowledge is another critical direction. For instance, hybrid architecture that combine graph neural networks for spatial relationships and RNNs for temporal dynamics could offer significant improvements in modeling complex building systems. Modular and hierarchical designs, which allow for the decomposition of large systems into manageable components, could further enhance scalability and interpretability.

*4.3.4. Improved training and optimization techniques*

The development of more efficient training and optimization algorithms is essential for advancing PIML. Current methods often struggle with high computational costs and hyperparameters tuning, which significantly affects convergence and performance, particularly for large-scale or real-time applications. Techniques such as multi-task learning, where models simultaneously optimize for multiple objectives (e.g., accuracy and physical consistency), automatic hyperparameter tuning based on rule importance evaluation [132] and advanced regularization methods such as physically hard-constrained methods [107,101] to enforce physical laws, could significantly improve both model efficiency and reliability. Additionally, the incorporation of uncertainty quantification in training could help address the inherent variability in building performance data and provide more robust predictions.

*4.3.5. Expansion to emerging applications*

While much of the focus has been on traditional building applications like IEQ assessment, energy modeling, control optimization and airflow simulation, future research could explore the potential of PIML in emerging areas such as renewable energy integration, smart grid optimization, indoor farming systems design, and urban-scale environmental simulations. However, the current state of PIML lacks a standardized schema for communication and interaction. For example, integrating an indoor airflow model with a thermal dynamic model could not only predict the assumed well-mixed conditions but also provide detailed temperature distributions, offering deeper insights for fine-grained control optimization. Addressing these limitations requires the development of a more general PIML modular framework for multi-domain, multi-scale, multi-component BPS. Such a framework should provide a user-friendly, plug-and-play interface to facilitate seamless extension and integration across various PIML applications.

*4.3.6. Standardization and benchmarking*

Table A.

To facilitate broader adoption, the development of standardized datasets, benchmarking protocols, and open-source tools is critical. Table B (appendix) illustrates an example of how PIML can be benchmarked in control and system optimization. It covers aspects such as modeling tasks, datasets (including building type, scope, climate zone, data source, HVAC system, input features, and data size), experiments (regarding control objectives, algorithms, prediction horizons, control horizons, and testbeds), and evaluation criteria. As summarized in Table B, these parameters vary from case to case, highlighting the need for a more standardized approach to benchmarking PIML. FAIR (Findable, Accessible, Interoperable, and Reusable) datasets [133] that cover diverse building types, climates, and operational scenarios could provide a common foundation for comparing PIML model performance. Similarly, standardized benchmarks with well-defined metrics—not only for evaluating accuracy but also for assessing scalability, data efficiency, computation time, and physical consistency—are critical for ensuring that advancements in PIML are rigorously validated and reproducible.

*4.3.7. PIML Embedded Foundation Models*

A foundation model is defined as "any model that is trained on broad data that can be adapted (e.g., fine-tuned) to a wide range of downstream tasks" [134]. Example include large language models (LLMs) such as OpenAI's GPT series, Google's BERT, which extend across carious modalities, including models like LLaMA, SALL-E. By fine-tuning with domain-specific knowledge and data, these models have shown great potential in building energy applications, such as energy modeling, data analysis, in-context learning, and knowledge extraction [135]. However, the effectiveness, efficiency, and explainability of foundation models are not guaranteed, particularly when addressing physical characteristics. Embedding PIML into foundation models offers a promising approach to developing BPS-specific, task-oriented foundation models. By leveraging the physics guarantees provided by PIML and the generalizability of foundation models, these hybrid approaches could address the unique challenges of building performance simulation while enhancing both accuracy and scalability.

**5. Conclusions**

PIML offers a transformative approach for BPS, effectively bridging the gap between traditional physics-based models and data-driven methods. By integrating physical principles into machine learning frameworks, PIML enhances predictive accuracy, generalization ability, and physical consistency—addressing the limitations of purely physics-based or data-driven approaches.

This review systematically examines the current state of PIML in BPS, defining the concept, summarizing key methodologies, and highlighting its applications in areas such as IEQ assessment, control and system optimization, and airflow simulation. The study also identifies existing research barriers, including challenges in balancing physics and data-driven components, integrating physical priors efficiently, and verifying model performance in real-world scenarios. Additionally, we highlight opportunities to advance PIML through adaptive methodologies, advanced neural architectures, and standardized benchmarks.

Looking forward, PIML has the potential to reshape BPS by enabling scalable, efficient, and physically consistent solutions for diverse applications, including smart building control, urban energy planning, and climate resilience enhancement for buildings. However, realizing this potential will require continued research to overcome current challenges, refine methodologies, and expand its applications. With a collaborative effort from researchers and practitioners, PIML can play a pivotal role in driving innovation and sustainability in the built environment.

**CRediT authorship contribution statement**

**Zixin Jiang:** Writing – review & editing, Writing – original draft, Visualization, Software, Resources, Methodology, Investigation, Conceptualization. **Xuezheng Wang:** Writing – review & editing, Writing – original draft, Methodology, Investigation, Conceptualization. **Han Li:** Writing – original draft, Methodology, Conceptualization. **Tianzhen Hong:** Writing – review & editing, Methodology, Conceptualization. **Fengqi You:** Writing – original draft, Methodology, Conceptualization. **Ján Drgoňa:** Writing – review & editing, Methodology, Conceptualization. **Draguna Vrabie:** Writing – review & editing, Methodology, Conceptualization. **Bing Dong:** Writing – review & editing, Writing – original draft, Supervision, Resources, Methodology, Funding acquisition, Conceptualization.

**Declaration of competing interest**

The authors declare that they have no known competing financial interests or personal relationships that could have appeared to influence the work reported in this paper.

The author is an Editorial Board Member/Editor-in-Chief/Associate Editor/Guest Editor for this journal and was not involved in the editorial review or the decision to publish this article.

**Acknowledgement**

Zixin Jiang, Xuezheng Wang and Bing Bong's work was supported by the U.S. National Science Foundation (Award No. 1949372). Han Li and Tianzhen Hong's work was supported by the Assistant Secretary for Energy Efficiency and Renewable Energy, Office of Building Technologies of the United States Department of Energy, under Contract No. DE-AC02-05CH11231. Jan Drgona's work was supported by the Ralph S. O'Connor Sustainable Energy Institute (ROSEI) at Johns Hopkins University (JHU). Draguna Vrabie's contributions were supported as part of the "Advancing Market-Ready Building Energy Management by Cost-Effective Differentiable Predictive Control" project by the U.S. Department of Energy, Office of Energy Efficiency and Renewable Energy, Building Technologies Office. PNNL is a multi-program national laboratory operated by Battelle Memorial Institute for the U.S. Department of Energy (DOE) under Contract No. DE-AC05-76RL0-1830.

**Appendix**

A. Literature review method

To conduct the review, we identified relevant articles published between 2010 and 2025 based on the inclusion criteria shown in Table A, using a two-round selection method. In the first round, the keywords were decomposed into three core concepts: "physics," "machine learning," and "building." These were then expanded to include similar terms—for example, ("physics" OR "physical" OR "physically"). Based on these keyword combinations, an initial search was conducted across major scientific databases, including Web of Science, ScienceDirect, Scopus, IEEE Xplore, Google Scholar, and SpringerLink. After the first-round selection, 219 papers remained. In the second round, we manually conducted a thorough review to remove duplicates and exclude articles unrelated to the topic. Finally, 50 papers most relevant to physics-informed machine learning in the building performance simulation domain were selected for in-depth analysis.

**Table A**
Search list for literature review.

| Category | Keywords |
| --- | --- |
| Paper type | Article, Conference |
| Language | English |
| Search period | 2010–01 to 2025–04 |
| Database | Web of Science, ScienceDirect, Scopus, IEEE Xplore, Google Scholar, and SpringerLink |
| Search terms | ("physics" OR "physical" OR "physically") AND ("neural network" OR "deep learning" OR "machine learning" OR "data-driven") AND ("building" OR "HVAC" OR "comfort" OR "CFD" OR "air") |

**Table B**
An example of PIML benchmark for control and system optimization.

| Application | Control and System Optimization | | | |
| --- | --- | --- | --- | --- |
| Modeling Task | Temperature [16,56–58,67–74,76] | Humidity [16,57,69] | $CO_2$ [68,70,71] | |
| Data Set | Building Type | Office [16,56,58,68–72,76] | Residential [57,66,67,73,74] | |
| | Scope | Single Zone [16,56,67,68,70,71,73,74] | Multi Zone [16,57,58,66,69,72,76] | |
| | Climate Zone | Tropical climatic [58,72] | Cool and humid [16,57,68–71] | |
| | Data Source | Experiment [16,56,58,67,71,76] | Simulation [57,66,68–70,72,74] | |
| | HVAC System | AHU [16,68–71,76] | Radiation Panel [67,76] | Heat Pump [57,73,74] |

**Table B** (*continued*)

| Application | | Control and System Optimization | | | |
|---|---|---|---|---|---|
| | Input Features | Outdoor Air Temperature [16,57,58, 66–74,76], | | | |
| | | Indoor Air Temperature [16,57,58, 66–71,73,74,76] | | | |
| | | Solar Radiation [16,57,58,67–73] | | | |
| | | Time [16,58,72,74,76] | | | |
| | | HVAC Power [16,57,58,66–71,73,74, 76] | | | |
| | | Adjacent Temperature [57,58,67,69,76] | | | |
| | | Occupancy [16,67,68,70,71,76] | | | |
| | | Relative Humidity [16,57,69] | | | |
| | | CO2 [68,70,71] | | | |
| | | Meta data [56,72] | | | |
| | | Setpoint [72] | | | |
| Experiment | Data Size | ≤ one week: [16,56,74] | ≤ one month: [16,74] | ≤ three months: [58,68, 70,73,76] | > six months: [67, 73,71] |
| | Obj. Function | Energy saving [67–74,76] | Thermal Comfort [16,66–74,76] | Flexibility [16,56–58] | |
| | Algorithm | MPC [57,58,67–72] | RL [71,73,74,76] | DPC [16,66,71] | RB [56] |
| | Prediction Horizon | ≤ one hour: [72] | ≤ six hours: [56,68,70,71,74,76] | ≤ one day: [16,57,58,67, 74] | > one day: [16,69,73] |
| | Control Horizon | ≤ five minutes: [58] | ≤ 15 min: [16,57,68–72,76] | ≤ one hour: [67,73,74] | |
| | Testbed | Virtual Testbed: [16,57,58,69,72–74] | Real World Implementation: [67,71,76] | | |
| Evaluation | Accuracy [16,56–58,68–70,73,74,76] | Consistency [16,57,58,68–70,73,76] | Computation [16,69,72] | | |

## Data availability

No data was used for the research described in the article.